\documentclass[12pt,letterpaper]{article}
\pdfoutput=1
\usepackage{graphicx}
\usepackage{subfig}
\usepackage{booktabs}
\usepackage{amsfonts,amssymb,amsmath}
\newcommand{\be}{\begin{equation}}
\newcommand{\ee}{\end{equation}}
\newcommand{\ben}{\begin{displaymath}}
\newcommand{\een}{\end{displaymath}}
\newcommand{\bea}{\begin{eqnarray}}
\newcommand{\eea}{\end{eqnarray}}
\newcommand{\bean}{\begin{eqnarray*}}
\newcommand{\eean}{\end{eqnarray*}}

\newcommand{\bra}[1]{\mbox{$\langle #1 |$}}
\newcommand{\ket}[1]{\mbox{$| #1 \rangle$}}

\newcommand{\eg}{{\it e.g.}}
\newcommand{\ie}{{\it i.e.}}

\newcommand{\tr}{\mbox{Tr}}

\newcommand{\commentout}[1]{}

\numberwithin{equation}{section}

\newcommand{\beq}{\begin{equation}}
\newcommand{\eeq}{\end{equation}}
\newcommand{\beqr}{\begin{displaymath}}
\newcommand{\eeqr}{\end{displaymath}}
\newcommand{\beqa}{\begin{eqnarray}}
\newcommand{\eeqa}{\end{eqnarray}}
\newcommand{\beqar}{\begin{eqnarray*}}
\newcommand{\eeqar}{\end{eqnarray*}}

\newcommand{\cO}{{\cal O}}

\newcommand{\non}{\nonumber}

\newcommand{\half}{\ensuremath{\frac{1}{2}}}

\newcommand{\ba}{\ensuremath{\bar{a}}}

\newcommand{\acosh}{\ensuremath{\mbox{arccosh}}}

\newcommand{\tth}{\ensuremath{t_{th}}}
\newcommand{\PK}[1]{\ensuremath{\hat{\mathbb{P}}_{#1}}}
\newcommand{\Kr}[1]{\mathcal{K}_{#1}}
\newcommand{\norm}[1]{\left\Vert#1\right\Vert}

\def\be{\begin{equation}}
\def\ee{\end{equation}}
\def\ba{\begin{eqnarray}}
\def\ea{\end{eqnarray}}

\def\tH{\widetilde{H}}

\def\tE{\widetilde{E}}
\def\bbeta{\bar{\beta}}
\def\NK{n}
\def\CH{H'}
\def\Ct{t'}

\begin{document}
%%%%%%%%%%%%%%%%%%%%%%%%%%%%%%%%%%%%%%%%%%%%%%%%%%%%%%%%%%%%%%%%%%%%%%%%
%%%%%%%%%%%%%%%%%%%%%%%%%%%%%%%%%%%%%%%%%%%%%%%%%%%%%%%%%%%%%%%%%%%%%%%%
%%%%%%%%%%%%%%%%%%%%%% TITLEPAGE %%%%%%%%%%%%%%%%%%%%%%%%%%%%%%%%%%%%%%%
%%%%%%%%%%%%%%%%%%%%%%%%%%%%%%%%%%%%%%%%%%%%%%%%%%%%%%%%%%%%%%%%%%%%%%%%
%%%%%%%%%%%%%%%%%%%%%%%%%%%%%%%%%%%%%%%%%%%%%%%%%%%%%%%%%%%%%%%%%%%%%%%%

\title{\LARGE \bf Thermalization of isolated quantum systems}

\author{S. Khlebnikov and M. Kruczenski
\thanks{E-mail: \texttt{skhleb@purdue.edu, markru@purdue.edu}}\\
        Department of Physics and Astronomy, Purdue University,  \\
        525 Northwestern Avenue, W. Lafayette, IN 47907-2036. }

\maketitle

\vspace{-1cm}

\begin{abstract}
 Understanding the evolution towards thermal equilibrium of an isolated quantum system is at 
the foundation of statistical mechanics and a subject of interest in such diverse areas as cold atom physics or the quantum 
mechanics of black holes. Since a pure state can never evolve into a thermal density matrix, 
the Eigenstate Thermalization Hypothesis (ETH) has been put forward by Deutsch and Srednicki as 
a way to explain this apparent thermalization, similarly to what the ergodic theorem does in classical mechanics. 
In this paper this hypothesis is tested numerically. First, it is observed that thermalization happens in a subspace of 
states (the Krylov subspace) with dimension much smaller than that of the total Hilbert space. 
We check numerically the validity of ETH in such a subspace, for a system of hard core bosons on a 
two-dimensional lattice. We then discuss how well the eigenstates of the Hamiltonian projected on the Krylov
subspace represent the true eigenstates. This discussion is aided by bringing the projected Hamiltonian to the
tridiagonal form and interpreting it as an Anderson localization problem for a finite one-dimensional
chain. We also consider thermalization of a subsystem and argue that generation 
of a large entanglement entropy can lead to a thermal density matrix for the subsystem well
before the whole system thermalizes. Finally, we comment on possible implications 
of ETH in quantum gravity.

\end{abstract}

\clearpage
\newpage

%\keywords{Thermalization}

%\preprint{\tt{} \\
%          \tt{hep-th/yymmnnn}  }

%%%% INTRODUCTION
\section{Introduction}
\label{intro}
 Statistical mechanics is based on the premise that a subsystem, weakly coupled to the rest of a large isolated 
system, eventually reaches a state of thermal equilibrium described by the canonical density matrix. This is the
density matrix that 
achieves maximum von Neumann entropy at fixed values of the average energy, number of particles 
and other additive conserved quantities. In principle there is no need to inquire about the precise state of the entire 
isolated system, as long as the latter can be assigned a sharply defined value of the total 
 energy (particle number, etc.). In particular, it does not have to be a thermostat with
 a large entropy or obey a certain (e.g., microcanonical) distribution. It can be in a single eigenstate of the total 
Hamiltonian, or in a pure state that is an arbitrary superposition of many such eigenstates with close-by energies.
 
 There is no doubt that this premise works extremely well in practice. On the other hand, when one attempts to justify 
it from first principles, one encounters the following question. Consider an initial state of the whole system
$\ket{\psi(t=0)}=\ket{\psi_0}$ with a narrow (in the sense made precise below) spread in energy $\Delta E$ around a mean value $E$.
 In the basis of energy eigenstates $\ket{E_\nu}$, the state, for any later time, is given by\footnote{We set $\hbar=1$ by measuring 
time in units of 1/Energy.}
 \beq
  \ket{\psi(t)} = \sum_\nu c_\nu e^{-i E_\nu t} \ket{E_\nu} \, ,
  \label{a1}
 \eeq
 where the $c_\nu$ are determined by the initial condition as $c_\nu=\bra{E_\nu}\psi_0\rangle$. 
Consider some observable $\hat{A}$ (for instance, the density of particles) pertaining 
to the subsystem in question. We expect it to thermalize, that is, its mean value in the 
state $\ket{\psi(t)}$ to reach, 
after a certain ``thermalization'' time,
a constant value independent of the initial state and given by the thermal expectation 
value, as obtained from the canonical density matrix of the subsystem. On the other hand,
the exact evolution of the mean value as a function of time is
 \beqa
  \bra{\psi(t)} \hat{A} \ket{\psi(t)} &=& \sum_{\nu,\nu'} c^*_{\nu'} c_\nu e^{-i (E_\nu-E_{\nu'}) t} \ \bra{E_{\nu'}} \hat{A} \ket{E_\nu}  \label{a2} \\
= \sum_{\nu} |c_\nu|^2 \ \bra{E_{\nu}} \hat{A} \ket{E_\nu} 
& + &
\sum_{\nu\neq \nu'} c^*_{\nu'} c_\nu e^{- i  (E_\nu-E_{\nu'}) t} \ \bra{E_{\nu'}} \hat{A} \ket{E_\nu} \, . \label{a4}
 \eeqa  
 After the thermalization time $\tth$, the last term, which contains the entire time-dependence, should reduce to a constant, up to small fluctuations.
The only way that can happen for a general initial state is if the individual off-diagonal matrix 
elements $\bra{E_{\nu'}} \hat{A} \ket{E_\nu}$ are small.\footnote{
Indeed, consider an initial state such that $c_\nu$ are nonzero 
(and of the same order) only for two values
of $\nu$, say, $\nu_1$ and $\nu_2$. Then, the time-dependent term in (\ref{a4}) is of order
$\bra{E_{\nu_2}} \hat{A} \ket{E_{\nu_1}}$ and is small only if that matrix element is small.}
The out-of-equilibrium initial state
is such that many off-diagonal terms add up coherently to give a sizable contribution. However, after the thermalization time $\tth$ the off-diagonal 
matrix elements no longer add up coherently and the second term in eq.(\ref{a4}) gives a small fluctuating contribution.  
Thus, the mean value of $\hat{A}$ becomes  
 \beq
 \left. \bra{\psi(t)} \hat{A} \ket{\psi(t)} \right|_{t\gg \tth} \simeq \sum_{\nu} |c_\nu|^2 \ \bra{E_{\nu}} \hat{A} \ket{E_\nu} \, , \label{a5}
 \eeq
 which is time-independent, in agreement with our a priori notion of thermal equilibrium. 
As it stands, however, the purported equilibrium 
value (\ref{a5}) seems to be strongly dependent on the initial state, 
namely, the expansion coefficients $c_\nu$. This seems to contradict the idea that it is
given by the thermal value.
 
 Thus, if we require that every possible initial state thermalizes, we are led to the Eigenstate Thermalization Hypothesis (ETH), 
put forward by Deutsch \cite{Deutsch} and Srednicki  \cite{Srednicki}. It states that, for those operators that thermalize,  the matrices in the basis of  
energy eigenstates have the property that their diagonal elements are smooth functions of energy:
 \beq
  \bra{E_{\nu}} \hat{A} \ket{E_\nu} = A(E_\nu) \, , \label{a6}
 \eeq
  while the off-diagonal elements are small enough, so that, at thermal equilibrium, they do not contribute significantly 
to any physical quantity of interest. 
 Now we can define a narrow band of energy $\Delta E$ such that the spread 
 $\partial_E A(E) \Delta E$ is of the same order as, or smaller than, the fluctuating contribution coming from the second term in eq.(\ref{a4}).  
The final result is that 
 \beq
 \left. \bra{\psi(t)} \hat{A} \ket{\psi(t)} \right|_{t\gg \tth} \simeq \sum_{\nu} |c_\nu|^2  A(E_\nu) \simeq A(E) \, ,
 \label{a7}
 \eeq
 independently of the initial state. Therefore ETH implies thermalization for every possible initial state, including the
energy
eigenstates, for which $\Delta E=0$. This has the interesting implication that individual energy eigenstates display thermal
 behavior.

ETH is rather nontrivial to check, because it requires diagonalization of the Hamiltonian 
matrix for a large system. Recently, some progress in this direction has been reported in 
\cite{ETHn, ETHn1, ETHn2,ETHn3}.

One of the primary motivations for the renewed interest in this topic
has been the level of isolation and control achieved in experiments with cold atoms 
\cite{therm}. Some of the recent developments in this area are described in the 
review \cite{reviews}. Single-eigenstate thermalization may be also relevant
to quantum computing, where thermalization due
to interactions among the qubits sets operational limits even if the whole system is
perfectly isolated \cite{QCapp}. Finally, a yet another, perhaps less expected,
area where understanding the precise mechanism of thermalization has become 
important is the physics of black holes, in particular, the properties of the Hawking 
radiation (see \eg\ \cite{AMPS} for a recent discussion). 
Notably, the AdS/CFT correspondence \cite{malda,string} 
relates thermalization of a quantum system to the formation
of black holes in quantum gravity, a process for which there is no clear theoretical 
description. 
 
As compared to the full ETH hypothesis, thermalization, 
in the sense of local observables reaching steady values, is much easier to test. 
This is because to evolve the
system numerically to the thermalization time $\tth$ requires access only to a very 
small subspace of the full Hilbert space---the Krylov subspace, generated by
repeated applications of the Hamiltonian to the initial state. Indeed, suppose first that
the initial state $\ket{\psi(t=0)}=\ket{\psi_0}$ 
%spreads strictly 
has support over an energy band of finite half-bandwidth $W = \half (E_{\max} - E_{\min})$, namely
\beq
 \ket{\psi_0} = \sum_{E_\nu=E_{\min}}^{E_{\max}}  \ket{E_\nu}\bra{E_\nu}\psi_0\rangle
\eeq
 This is always the case if the full spectrum is bounded from above and  below, as is certainly true
for the system of a finite number of bosons on a lattice that we study in this paper. The state at arbitrary time $t$ is given by 
\beq
\ket{\psi(t)} = e^{-i H t} \ket{\psi_0} = 
\sum_{p=0}^\infty \frac{(-i)^p}{p!} t^p H^p\ket{\psi_0} \, . \label{b3}
\eeq
By assumption, the energy is bounded from below and above, implying that the series is absolutely 
convergent and therefore, to any finite precision required in the calculations, 
can be truncated at a finite number ($\NK$) of terms, 
showing that only an appropriately chosen Krylov subspace 
\beq
\Kr{\NK}=\mbox{span}\{H^p\ket{\psi_0}, p=0\ldots \NK - 1 \}   \, ,     \label{b4}
\eeq 
is required to follow the evolution. For fixed precision, the larger the time, 
the larger the dimension of the Krylov subspace that needs to be considered. To reach
$t=\tth$, however, it is typically sufficient to consider only an $\NK$ 
vastly smaller than the full dimension $N$ 
of the Hilbert space.
 
Our conclusion then is that it is a useful approximation to replace
(\ref{b3}) with
\beq
 \ket{\psi(t)} = e^{-i H t} \ket{\psi_0} \simeq \PK{\NK} e^{-i H t} \ket{\psi_0}   \, , \label{b5}
\eeq
where $\PK{\NK}$ projects onto a Krylov subspace of dimension $\NK$ with the 
base state $\ket{\psi_0}$. We have used this method to study thermalization of systems of
hard core bosons on 2-dimensional lattices. We have considered onset of the thermal behavior 
for two types of quantities. One is the average occupation numbers of various sites of the lattice 
or, alternatively, of various single-particle modes; the other is the entanglement
(von Neumann) entropy of a subsystem.\footnote{The initial state of $\ket{\psi_0}$ of 
the entire system is a pure state, and will remain such upon evolution. Thus, the
von Neumann entropy of the entire isolated system is zero.}
 
Our results can be used for a partial check of the ETH. The approximate evolution 
equation (\ref{b5}) can 
be cast in a form similar to (\ref{a1}):
\beq
 \ket{\psi(t)} \simeq \sum_\ell \tilde{c}_\ell e^{-i\tE_\ell t}\, \ket{\tE_\ell}
\eeq
 The only difference is that, 
instead of the eigenstates $\ket{E_\nu}$ of $H$, we are now using the Ritz vectors 
$\ket{\tE_\ell}$, namely, the eigenstates 
of the projected (or reduced) Hamiltonian
\be
\tH = \PK{\NK} H \PK{\NK} \, .
\label{PKH}
\ee
 We will argue that a Ritz vector $|\tE_\ell\rangle$ 
contains,
with significant amplitudes, only those eigenstates of $H$ that fall into a narrow
band of energies around $\tE_\ell$. The width 
$\Delta E_\ell = \bra{\tE_\ell} (H-\tE_\ell)^2\ket{\tE_\ell}^{1/2}$ 
of this band scales as $1/\sqrt{n}$ with
the dimension $n$ of the Krylov subspace. This allows us to estimate the expectation
values of the operator $\hat{A}$ in the Ritz states,
\be
\bra{\tE_{\ell}} \hat{A} \ket{\tE_\ell} \equiv \widetilde{A}(\tE_\ell) \, ,
\label{KETH}
\ee
as follows. Suppose that, for the
exact energy eigenstates $\ket{E_\nu}$
with energies near $\tE_\ell$, the diagonal element (\ref{a6}) is 
a smooth function of energy, and the off-diagonal elements are negligible. Then, 
\be
\widetilde{A}(\tE_\ell) = \sum_\nu |c_{\ell\nu}|^2 A(E_\nu) \, ,
\label{ARitz}
\ee
where $c_{\ell \nu} = \langle E_\nu | \tE_\ell \rangle$. 
Since $|c_{\ell\nu}|^2$ is a sharply peaked
function of $E_\nu$, with average energy $\langle E_\nu \rangle = \tE_l$ and standard 
deviation $\Delta E_\ell$,
and $A(E_\nu)$ is smooth (over a much broader range of energies), we can estimate
(\ref{ARitz}) by using the Taylor expansion for $A(E_\nu)$ near $E_\nu = \tE_\ell$. The
result is
\be
\widetilde{A}(\tE_\ell) = A(E_\nu) + \half A''(E_\nu) (\Delta E_\ell)^2 + \ldots
=  A(E_\nu) + O(1/n) \, .
\label{estRitz}
\ee
Thus, under the stated conditions, $\widetilde{A}(\tE_\ell)$ 
is a smooth function of $\tE_\ell$, at least up to 
$O(1/n)$ corrections. Our numerical results suggest that, for a large system,
$\widetilde{A}(\tE_\ell)$ may in fact be 
smooth to an accuracy better than $O(1/n)$. That would imply that
the $O(1/n)$ correction in (\ref{estRitz}) is also a smooth function of $\tE_\ell$. 
In particular, we will confirm numerically that $\Delta E_\ell$ is smooth.

 Summarizing, if we were to find that $\widetilde{A}(\tE_\ell)$ does not become smoother as we increase the size of the full 
Hilbert space then ETH does not hold. 
 The converse is not true: it is possible to imagine that $A(E_\nu)$ has large fluctuations that average out
when constructing the Ritz vector and resulting in a function  $\widetilde{A}(\tE_\ell)$ smoother than $A(E_\nu)$. In 
that case, the Taylor expansion leading to eq.(\ref{estRitz}) is not valid, and the two functions are not directly related.

% Summarizing, testing ETH in the Krylov subspace can invalidate ETH but not prove its validity in the whole space. 
% Equivalently, the validity of ETH in the Krylov subspace implies thermalization only for those initial state that are 
% contained in such a subspace.

\section{Summary of results}

\begin{figure}
\centering
\includegraphics[width=10cm]{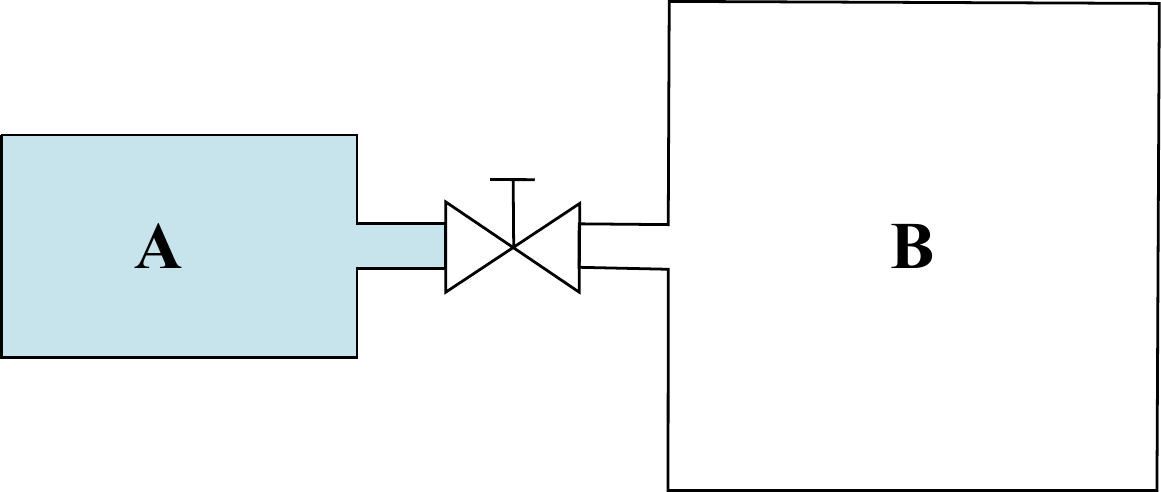}
\caption{A schematic system for modeling expansion of a gas into a larger container. 
In this paper we consider a lattice version of such a system
and follow its quantum evolution numerically.}
\label{system}
\end{figure}

 We consider a two-dimensional lattice gas of bosons with hard-core
repulsion and an additional nearest-neighbor repulsive interaction. 
Similarly to \cite{ETHn}, the system, schematically depicted in Fig.~\ref{system}, is spatially separated into two regions 
(``boxes'') of different sizes and we study numerically the expansion of the gas, originally
in the smaller box, into the larger one. 
Unlike ref.~\cite{ETHn}, we do not diagonalize the
full Hamiltonian but instead follow the evolution in the Krylov subspace (\ref{b4}).
This gives us access to much larger systems. 
Numerically, we have considered systems with dimension $N$ of the Hilbert space up to 
$N\simeq 10^7$. We have found that the dimension $n$ of the Krylov subspace required to follow 
the evolution up to and somewhat beyond thermalization depends only on such time and is typically 
of order of a few thousand, $\NK \simeq 10^3$ independently of the size $N$ of the full Hilbert space.  

In this setup, one can consider thermalization of the gas either by
explicitly comparing the properties of the gas to those of a thermal state, or by
verifying the Eigenstate Thermalization Hypothesis (ETH) as it applies to the Krylov
subspace; we refer to this as the ``Krylov ETH'' (KETH). As discussed in the introduction,
verification of the KETH provides a partial check of the ETH in the full Hilbert space.

 Numerically, verification of the KETH amounts to diagonalizing the reduced  Hamiltonian (\ref{PKH}) 
and verifying the two statements of the hypothesis: (i) the smoothness, as a function of energy, of the diagonal
elements (\ref{KETH}) of a suitable operator $\hat{A}$ and (ii) the smallness of the
off-diagonal elements. Here we present results for $\hat{A} = n_i$, the occupation number of lattice site $i$ and $\hat{A}=n_k$ the single particle states occupation number.
%For hard-core bosons, $n_i^2 = n_i$, which allows one to obtain an estimate of %the 
%off-diagonal matrix elements of this operator as follows. Write a 
%diagonal element (in an eigenstate of the full Hamiltonian) as
%\beq
%  \bra{E_\nu} n_j \ket{E_\nu} =\bra{E_\nu} n_j^2 \ket{E_\nu} = \sum_{\nu'} %|\bra{E_\nu} n_j \ket{E_{\nu'}}|^2  
%\le 1   \, .    \label{b15}
%\eeq 
%Assuming that the sum over $\nu'$ contains order $N$ terms of comparable %magnitude
%($N$ is the dimension of the
%full Hilbert space), we conclude that the typical value of an off-diagonal %matrix
%element is $|\bra{E_\nu} n_j \ket{E_{\nu'}}| \sim \frac{1}{\sqrt{N}}$, as %required by the ETH
%\cite{Srednicki}. 
%This 
%is sufficient to justify
%the neglect, at large enough times, of the time-dependent term in (\ref{a4}). 
 
 Regarding the smoothness of the diagonal elements as a function of energy, the results can be seen in Figs.\ref{KrylovETH} and \ref{KrylovETHnk}. 
It is apparent that, as $N$ is increased, the curves become smoother except at the edges of the spectrum. 
 It is well-known that, even for moderate $\NK$, say $\NK \sim 10^3$, the Krylov subspace 
methods find some eigenstates of the full problem essentially exactly:
\be
\ket{\tE_\ell} = \ket{E_\nu} \, .
\label{exact}
\ee
These are precisely the eigenstates corresponding to the eigenvalues near the bottom and top of the spectrum.
For these eigenstates, testing the KETH is equivalent to testing the full ETH.
As mentioned, the diagonal matrix elements of $n_i$ or $n_k$ in these states 
are not particularly smooth, and do not get visibly
smoother as the dimension $N$ of the Hilbert space increases. These very low and very
high energy states, however, should probably not be expected to thermalize
in the first place, and so the
failure of the smoothness condition for these states does not indicate a failure
of the ETH. We will therefore pay most attention to the states in the middle of the 
spectrum. For those, the Ritz vectors represent (in the sense that will be made more
precise later) narrow bands of the true spectrum, rather than the individual eigenstates.
We find that, as we increase the total size of the Hilbert space $N$ at fixed $\NK$,
the matrix element of $n_i$ and $n_k$ in these states do become smoother, see Figs.~\ref{KrylovETH} and~\ref{KrylovETHnk}.
As noted in the introduction, this is consistent with the ETH, but does not prove it:
except at the edges of the spectrum,
the eigenstates of the projected Hamiltonian 
span a large number of exact energy eigenstates and therefore their properties average those of the exact eigenstates. For that reason the validity of KETH does not
necessary extend to the whole Hilbert space. Equivalently, we can say that we showed that the given initial state thermalized due to ETH in its Krylov subspace but we cannot
show that every state thermalizes. Although in practice we tried several other initial states and all thermalized, they are still a tiny fraction of all possible states.  

 Regarding the smallness of the off-diagonal elements, we plotted $\bra{\tilde{E}_\ell} n_i \ket{\tilde{E}_{\ell'}}$ as a function of $\tilde{E}_\ell-\tilde{E}_{\ell'}$
 in Fig.~\ref{off-diag} where it can be seen that the off-diagonal elements become smaller as the size $N$ of the Hilbert space increases. It should be noted that only matrix elements between Ritz states in the central region where $\bra{\tilde{E}_\ell} n_i \ket{\tilde{E}_{\ell}}$ is smooth are plotted. 
At the edges of the spectrum the off-diagonal elements are large as expected since those states do not thermalize. 

A convenient way to diagonalize (\ref{PKH}) is to first construct a special basis in
the Krylov subspace by using the Lanczos method \cite{Lanczos}. 
Namely, starting from the initial state $\ket{\psi_0}$ a 
basis of $\Kr{n}$ is constructed by successive application of $H$ and orthogonalization. 
A special property of the Lanczos basis is that the Hamiltonian in this basis is 
tridiagonal.\footnote{Numerically, full reorthogonalization every certain number
of steps is used to avoid accumulation of errors.} We use analytical estimates 
to argue that the off-diagonal elements of the Hamiltonian are 
approximately constant of value $\Delta E$ and the diagonal elements 
can be considered as if taken from a random 
distribution with dispersion $\sim\frac{\Delta E}{\sqrt{\Delta N}}$.  
Here $\Delta E$ is the range of energies associated with the Krylov subspace, and
$\Delta N$ is the number of states in this range. For a sufficiently large KS, we expect
$\Delta E\sim W$, the half-bandwidth of the system, and $\Delta N \sim N$, the full
Hilbert-space dimension. We find these estimates consistent with our numerical results.

This general form of the Hamiltonian in the Lanczos basis
is the same as that of a tight-binding 
Hamiltonian for a particle hopping on a one-dimensional chain with a random disorder
potential. This analogy allows us to apply results pertaining to Anderson localization
phenomena to understand some properties of the Ritz states in our case. In particular,
we can identify the states at the edges of the spectrum, for which Lanczos iterations
have already converged, with the localized states in the Anderson problem, 
and the states in the middle of the
band, for which $\bra{\tE_\ell} n_i\ket{\tE_\ell}$
varies smoothly,  with the extended states. 

The state of the entire isolated system in our computations is always a pure state. 
Initially, in fact, it is a product of a pure state of the subsystem corresponding
to the small box, and the vacuum of the rest. The subsystem, however, does not remain
in a pure state but quickly 
transitions to a mixed state, described by a density matrix.  
Let us denote the subsystem as $A$ and its density matrix 
(resulting from tracing over the rest of the system) as $\rho_A$. 
It is known that, for given mean values of the energy and 
particle number, the thermal density matrix $\rho_{th}$ is the one that has 
largest entropy as defined by
\beq
  S_A = - \tr [\rho_A \ln \rho_A ] \, .  \label{b1}
\eeq
This entropy is equal to the entanglement entropy between the two containers and can be computed numerically for our system. A priori, one expects the following behavior: $S_A$ starts
from zero (since the initial state is a product state), grows to some value
due to streaming of the particles into the larger box and, then, either stays near that
value, if the larger box is relatively small, or starts to decrease, as the gas leaves the container
and the number of available states decreases. Overall, this pattern of growth and decrease
is reminiscent of the behavior discussed by
Page \cite{Page} in his work on information in black holes.

The question we wish to answer here is to what 
extent, and when, does the density matrix of the subsystem become a thermal one.
The result is that, for the initial state we consider, the entropy increases
rapidly at the beginning (first as $S\sim - t^2 \ln (t/t_0)$ as can be shown analytically, 
and then approximately linearly) until it reaches a value close to the maximum allowed \ie\
the thermal one. 
Remarkably, this occurs early in the expansion; in other words, 
the subsystem reaches an actual thermal state essentially by streaming particles into 
the vacuum. After this thermalization of the subsystem, the entanglement entropy starts to decrease.
At first, the decrease is rapid,
as the particles continue to leave the small container. Later in the evolution, 
the large container fills up and begins to supply particles back to the small one. 
As a result, the entropy decreases 
more slowly until it reaches its final thermal value, corresponding to the equilibrium of the
entire system. The local thermal state corresponding to the maximum of $S_A$
is analogous to the thermal state of a black hole, and its subsequent decrease to a decrease of the
black-hole entropy by Hawking radiation. Existence of this state leads us to assign
a special significance to local variables, in the sense that they are the ones that would most
naturally obey the ETH. This is one of the reasons that
we concentrated on the local occupation number in our tests of the ETH. 

%After this summary of the main results let us proceed to describe the concrete %physical system 
%and the numerical methods by which they are obtained. 

\section{Physical system and numerical methods}

 The results of this paper are argued in a general manner but based on a concrete case where numerical methods allow us to follow the
quantum evolution of a system with precision limited only by round-off error, which 
is much smaller than the fluctuations due to the time-dependent term in (\ref{a4}). 
Having this degree of control comes at a price, 
as we can only consider small systems. The time evolution was restricted to systems 
such that the dimension of the Hilbert space $N \lesssim 10^8$. When constructing the Krylov subspace we were restricted to a Krylov 
subspace of order $10^3$ for a system whose Hilbert space has a dimension $\lesssim 1.5\times 10^7$. It is important to note that the system has no symmetries and therefore
it has no invariant subspaces. 

 The system we consider is similar to the one considered in \cite{ETHn} where the ETH hypothesis was tested by exact diagonalization in the full Hilbert space (i.e., without considering a Krylov subspace). That limited
the maximum dimension of the Hilbert space considered to $~2\times 10^4$, substantially smaller
than the one considered here. For that reason, the effects of the ETH hypothesis should be more evident in our case. In the works \cite{ETHn1,ETHn2,ETHn3} a one dimensional system was considered instead. The Hilbert
space of the systems considered there has dimension up to $\sim 10^6$ but the Hamiltonian is block diagonal with blocks of order up to $\sim 30,000$, allowing exact diagonalization. Since ETH suppresses fluctuations 
proportionally to the inverse of the square root of the number of states that the Hamiltonian mixes, in that case the relevant dimension should still be $\sim 3\times 10^4$. 

\begin{figure}
\centering
\includegraphics[width=10cm]{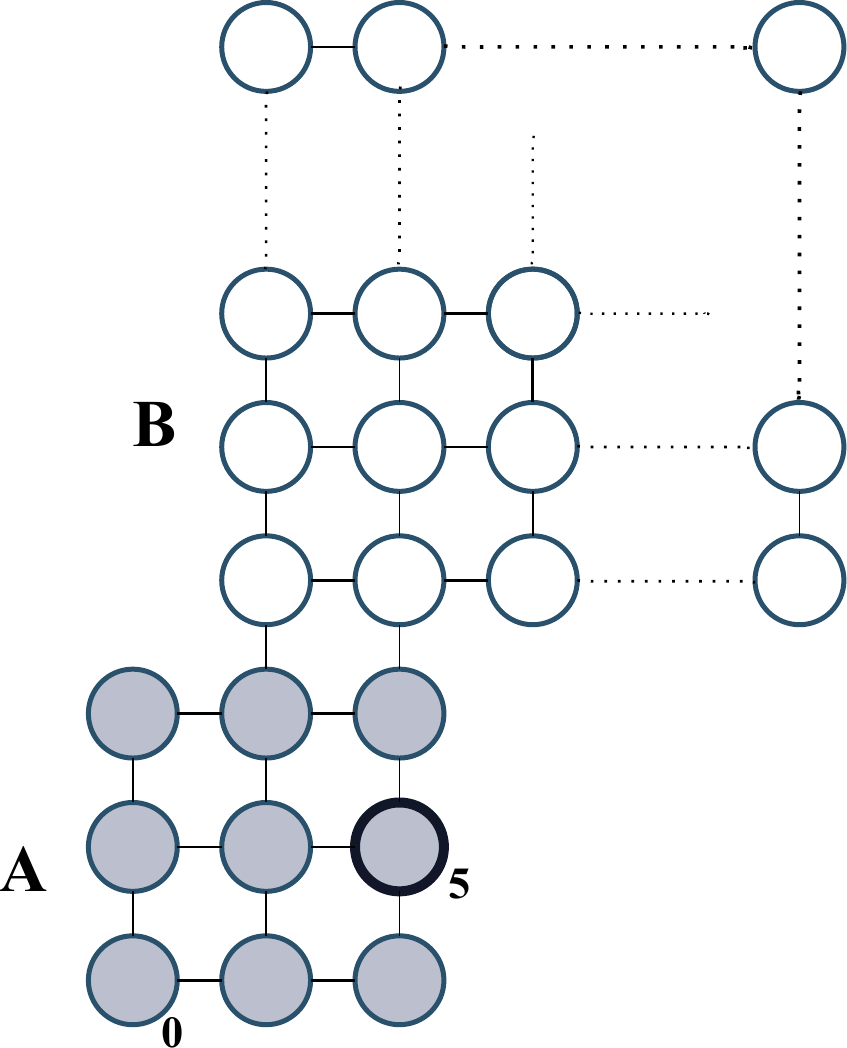}
\caption{Lattice considered in this paper. The bosons are initially confined to the $3\times 3$ shaded sublattice $A$, and subsequently expand into the larger 
sublattice $B$. The larger square $B$ has size $n_2\times n_2$ where we used values $n_2=4,5,6,8,10$. For reference, the sites are numbered from $0$ and increasing 
towards the right and up. Site $i=5$ is in bold since the occupation number on that site is depicted later in the paper. Site $i=20$ is also considered but its position depends on the size of sublattice $(B)$.  }
\label{Lattice}
\end{figure}

 Going back to the present work, the system studied here is a two dimensional lattice gas of hard core bosons that can hop between nearest 
neighbors of the lattice and have an additional near neighbor repulsion. 
 The Hamiltonian is
\beq
 H = J_1 \sum_{\langle i,j\rangle} (a_i a^\dagger_j + a^\dagger_i a_j) + 2 J_0 \sum_{\langle i,j\rangle} n_in_j \, ,  \label{b2}
\eeq
where the sum is over pairs of near neighbors and, for concreteness\footnote{We tried other values of the couplings but no substantial difference was observed.}, 
we take $J_1=-1$, $J_0=\frac{1}{4}$. The occupation numbers $n_i$ take values $n_i=0,1$ in view of the hard core property of the bosons. 
The shape of the lattice is depicted in fig.\ref{Lattice} and can be described as a small volume where all bosons are contained initially and a large volume into 
which the bosons expand. Therefore, this system describes the classic thermodynamic problem of the expansion of a gas into a larger container, the difference 
is that we follow the quantum state of the system exactly. To be precise, the precision is limited by machine precision (we mostly use long double, \ie\ 16 bytes 
real numbers). We tested that the evolution is completely reversible and also that the final state is the same when using a variety of different procedures 
to perform the time evolution. 

 The lattice sizes considered in this paper are shown in table \ref{LatticeSizes}.
% \begin{table}[h]
% \caption{Systems considered}
% \begin{tabular}{|ccc|r|r|r|}
% \hline
% \multicolumn{3}{|c|}{System} & \# of sites & \# of bosons & Dim. of Hilbert space \\ 
% \hline 
% $3\times 3$ &+& $5 \times 5$   &  34  & 3 & 5,984       \\
% $3\times 3$ &+& $4 \times 4$   &  25  & 5 & 53,130      \\
% $3\times 3$ &+& $6 \times 6$   &  45  & 5 & 1,221,759   \\
% $3\times 3$ &+& $8 \times 8$   &  73  & 5 & 15,020,334  \\
% $3\times 3$ &+& $10 \times 10$ & 109  & 5 & 116,828,271 \\
%\hline
%  \end{tabular}
%  \label{LatticeSizes}
%\end{table} 
\begin{table}[h]
 \centering
  \renewcommand{\arraystretch}{1.2}
 \begin{tabular}{@{}cccrrr@{}}
 \toprule
 \multicolumn{3}{c}{System} & \# of sites\ & \# of bosons\ & dim. of Hilbert space \\ 
 \midrule
 $3\times 3$ &+& $5 \times 5$   &  34  & 3 & 5,984       \\
 $3\times 3$ &+& $4 \times 4$   &  25  & 5 & 53,130      \\
 $3\times 3$ &+& $6 \times 6$   &  45  & 5 & 1,221,759   \\
 $3\times 3$ &+& $8 \times 8$   &  73  & 5 & 15,020,334  \\
 $3\times 3$ &+& $10 \times 10$ & 109  & 5 & 116,828,271 \\
\bottomrule
  \end{tabular}
  \caption{Systems considered}
  \label{LatticeSizes}
\end{table} 
 The first case is the only one that we can diagonalize exactly and therefore it is used as a test of the Krylov subspace methods. 

 The initial state of the system was taken to be an energy eigenstate of the Hamiltonian of the small $3\times 3$ container with 5 bosons and 
approximately in the middle of the spectrum (state $\nu=40$ if $\nu=0$ is the ground state). A similar state was chosen for the 3 bosons case.
 Other initial states including eigenstates of occupation number were considered but the results are not displayed here since they are substantially the same. 
 
 Given an initial state, the Krylov subspace is constructed using the Lanczos method\footnote{The numerical methods are described in general later in the paper,
here we only give the computational details as they pertain to our calculation.} 
 with full reorthogonalization. 
The Lanczos method increases the dimension of the Krylov subspace iteratively, in unit steps.
At each step the matrix of the occupation
numbers for certain sites is computed and stored in memory. The Lanczos vectors are stored in disk storage to be used for reorthogonalization every 10 or 20 steps. 
 At certain dimensions of the Krylov subspace, 
the tridiagonal matrix is diagonalized and the occupation numbers as functions of energy are saved. 

Time evolution in the Krylov subspace 
in principle does not require diagonalization of the projected Hamiltonian: a convenient 
method is based on expansion of the evolution operator $\exp(-i H t)$, where $H$ is the full Hamiltonian,
in Chebyshev polynomials of $H$ and applying these to the initial state directly. 
The primary method used in this paper was the Chebyshev polynomial expansion to order $20$ for time step\footnote{Since we take $\hbar=1$, the unit of 
 time is determined by choosing the constants $J_0$ and $J_1$ in the Hamiltonian.} $\Delta t=0.25$. 
As a check we evolved the system up to time $t=150$ (past thermalization) and checked that the resulting
state agreed to $10^{-14}$ precision with evolution in a single step, the latter using 
the Chebyshev polynomial expansion to order $\sim 3000$.  
 
 In selected cases we used the Ritz vectors (eigenvectors of the projected Hamiltonian) to perform the time evolution and found agreement with the Chebyshev expansion. 

 Most computations were done on a system with two 6-core Intel Xeon CPUs with 48GB RAM. Some computations were done using a node with four 12-core AMD Opteron CPUs and 
96GB RAM and, alternatively, in an NVIDIA Tesla 2700 GPU.

\section{Thermalization and ETH hypothesis, numerical results}
\begin{figure}
\centering
\includegraphics[width=10cm]{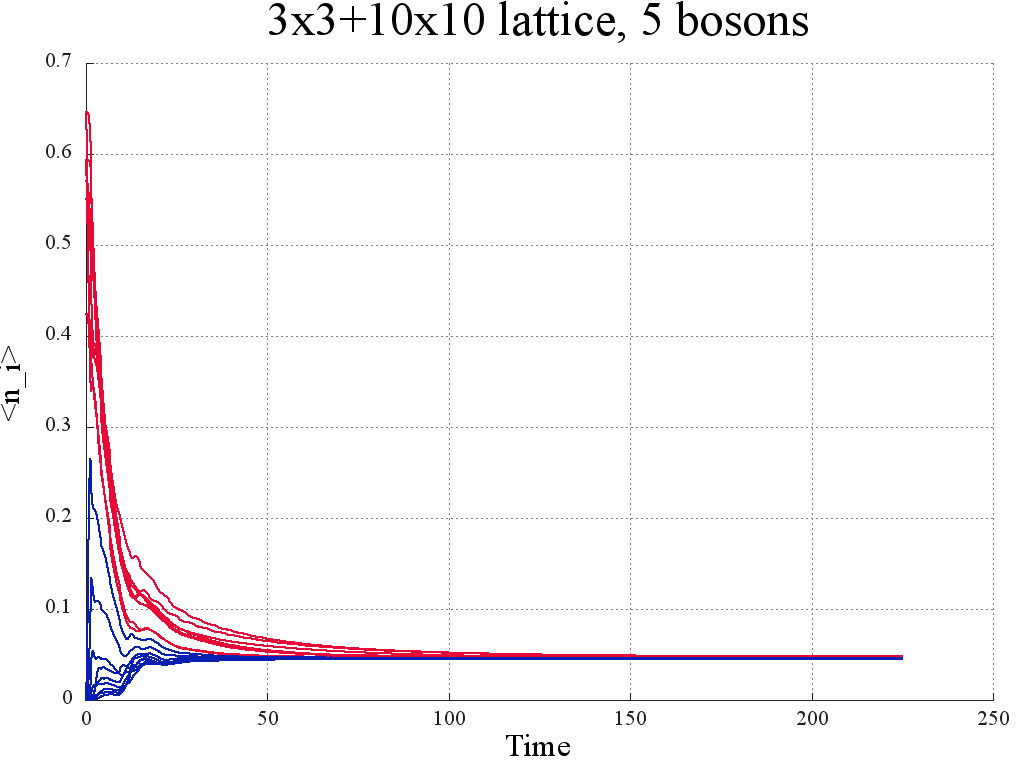}
\caption{Occupation numbers of different sites as functions of time. After the thermalization time they become approximately constant in time and equal to one another.
The red curves correspond to sites in region $A$ and the blue ones to sites in region $B$.}
\label{Thermalization}
\end{figure}

We follow the evolution of a quantum system in the manner described above and compute expectation values 
of certain operators, to see if they thermalize.
For the particular system under consideration, we have computed the mean values of the occupation numbers 
of all sites. These exhibit thermal behavior as shown in fig.\ref{Thermalization}, 
namely, after a certain time $\tth\sim 100$ the occupation numbers become approximately time independent and, in this particular system,
equal to each other. The time independence is clearly up to some fluctuations; 
these are a measure of the time-dependent term in (\ref{a4}). It should be noted that fluctuations in the mean value are small but quantum fluctuations
are large since any individual measurement of $n_i$ will give an integer. 

\subsection{Time evolution}

 To understand thermalization from a physical point of view, the dynamics needs to be followed only up to a precision of the order of  the time-dependent term 
in (\ref{a4}).
 The objective of this paper is of course to follow the system much more precisely, to be sure that the thermalization observed is due to the actual dynamics of the system and not to approximations made along the way. Nevertheless, there is still a finite precision that one can achieve numerically. For that reason, it is useful, both theoretically and
 practically, to discuss approximations to the time evolution of the system. 

The central approximation involved is replacing the exact evolution (\ref{b3}) with the projected 
evolution  (\ref{b5}). This is equivalent to retaining only a finite number of terms in the series
expansion of the evolution exponent (\ref{b3}). We have justified that by noting that, for 
a bandwidth-limited Hamiltonian, the series (\ref{b3}) is absolutely convergent and so a finite number of terms
is sufficient to follow the evolution to any desired accuracy.

 Instead of simply truncating the expansion of (\ref{b3}) it is numerically more stable to use one of two known alternative procedures \cite{numerics}. 
One is to diagonalize the matrix of the Hamiltonian in the Krylov subspace and use the eigenvalues 
and eigenvectors to compute the evolution. 
The eigenvectors $\ket{\tE_\ell}$ are the Ritz vectors, which satisfy
\beq
  \PK{n} H \ket{\tE_\ell} = \tE_\ell \ket{\tE_\ell}   \, .  \label{b6}
\eeq
 The approximate time evolution is given by
\beq
 e^{-iHt} \ket{\psi_0}\simeq \PK{n} e^{-iHt} \PK{n} \ket{\psi_0} = \sum_{\ell=0}^{n-1}  
e^{-i\tE_\ell t} \ket{\tE_\ell} \bra{\tE_\ell} \psi_0\rangle   \, , \label{b7}
\eeq
where we have used $\PK{n} \ket{\psi_0}=\ket{\psi_0}$ and $(\PK{n} H \PK{n})^m \ket{\psi_0} = \PK{n} H^m \ket{\psi_0}$, Notice that $ \PK{n} H^m \ket{\psi_0}= H^m \ket{\psi_0}$ if $m\le n$ and vanishes otherwise.
 The other method requires first to shift and rescale the Hamiltonian in such a way that the spectrum is in the interval $(-1,1)$. If the spectrum of $H$
is contained in the interval $(\bar{E}-W,\bar{E}+W)$ we define
\beq
\CH = \frac{1}{W} (H-\bar{E})   \, .   \label{b8}
\eeq 
  Defining $\Ct =t\, W$ we can use the expansion \cite{numerics}
\beq
  e^{-i \CH \Ct}\ket{\psi_0}= J_0(\Ct) \ket{\psi_0} + 2 \sum_{n=1}^{\infty} (-i)^n J_n(\Ct) T_n(\CH) \ket{\psi_0}  \, ,  \label{b9}
\eeq
where $T_n(\CH)$ is a Chebyshev polynomial and $J_n$ the Bessel functions. Since the Chebyshev polynomial $T_n$ has order $n$, $T_{n'\le n}(\CH)\ket{\psi_0}$ is in the Krylov subspace $\Kr{n}$. For that reason
\beqa
 \PK{n} e^{-i\CH \Ct} \ket{\psi_0} &=& J_0(\Ct) \ket{\psi_0} + 2 \sum_{n'=1}^{n} (-i)^{n'} J_{n'}(t) T_{n'}(\CH) \ket{\psi_0} \\
 && + 2 \PK{n} \sum_{n'=n+1}^{\infty} (-i)^{n'} J_{n'}(\Ct) T_{n'}(\CH) \ket{\psi_0}   \, . \label{b10}
\eeqa
In view of the behavior of the Bessel functions for large order, fixed argument \cite{SF}
\beq
 J_n(\Ct) \sim \frac{1}{n!} \left(\frac{\Ct}{2}\right)^n \sim \frac{1}{\sqrt{2\pi n}} \left(\frac{e\Ct}{2n}\right)^n  \, ,   \label{b11}
\eeq
the last sum can be discarded for values
\beq
 n \gg \frac{e \Ct}{2}   \, .   \label{b12}
\eeq 
 For large values of $t$ this formula overestimates the required expansion order $n$ since for $t$, $n$ large, even for $n\gtrsim\Ct$, 
 with fixed ratio $\Ct/n<1$ the Bessel function 
is exponentially small  for large $n$. The requisite asymptotics is \cite{GR,SF}
\beq
 J_n(\Ct) \simeq \frac{1}{\sqrt{2\pi}} \frac{1}{(n^2-\Ct^2)^{\frac{1}{4}}}\, \exp(\sqrt{n^2-\Ct^2}-n\,\acosh\frac{n}{\Ct}) \, ,    \label{b13}
\eeq
valid for $n\rightarrow \infty$, $\frac{\Ct}{n}<1$ fixed. Overall, 
\beq
 n \gtrsim \Ct = t\, W  \, ,   \label{b14}
\eeq
 is an appropriate estimate of the required order of the expansion. Summarizing, to follow the evolution of the system to a certain time $t$ we need to consider only 
a Krylov subspace of order $n\sim t\, W$. Beyond that, the terms decrease faster than exponentially in $n$. Numerically, we checked this in two different ways. 
For the smallest system we compared the evolution using the Chebyshev approximation with the evolution using the exact eigenstates and verified their equivalence to 
machine precision. For the systems that we cannot diagonalize exactly, the evolution was tested by evolving 
to time $t= 150 \frac{1}{W}$ in small steps 
$\Delta t=0.25 \frac{1}{W}$ 
requiring an expansion of order $n=20$ and in a single step of size $t$, requiring $n=2700$. The resulting vectors agree component by component 
 to a $10^{-14}$ precision. For that reason the Chebyshev method was used to compute the time evolution throughout this paper. 

\subsection{ETH in the Krylov subspace}
 
  Given eq.(\ref{b7}) it is clear that thermalization happens if ETH is valid in the Krylov subspace, namely, 
if the matrix of the relevant operator in the 
basis of Ritz vectors (eigenvectors of the projected $H$) is such that the diagonal elements are smooth functions of the energy and the off diagonal elements are small. 
So let us now test this numerically for some selected operators. 

\subsubsection{Site occupation numbers}
\begin{figure}
\centering
\subfloat[$N=5984, n=1240$    ]{\includegraphics[width=6.5cm]{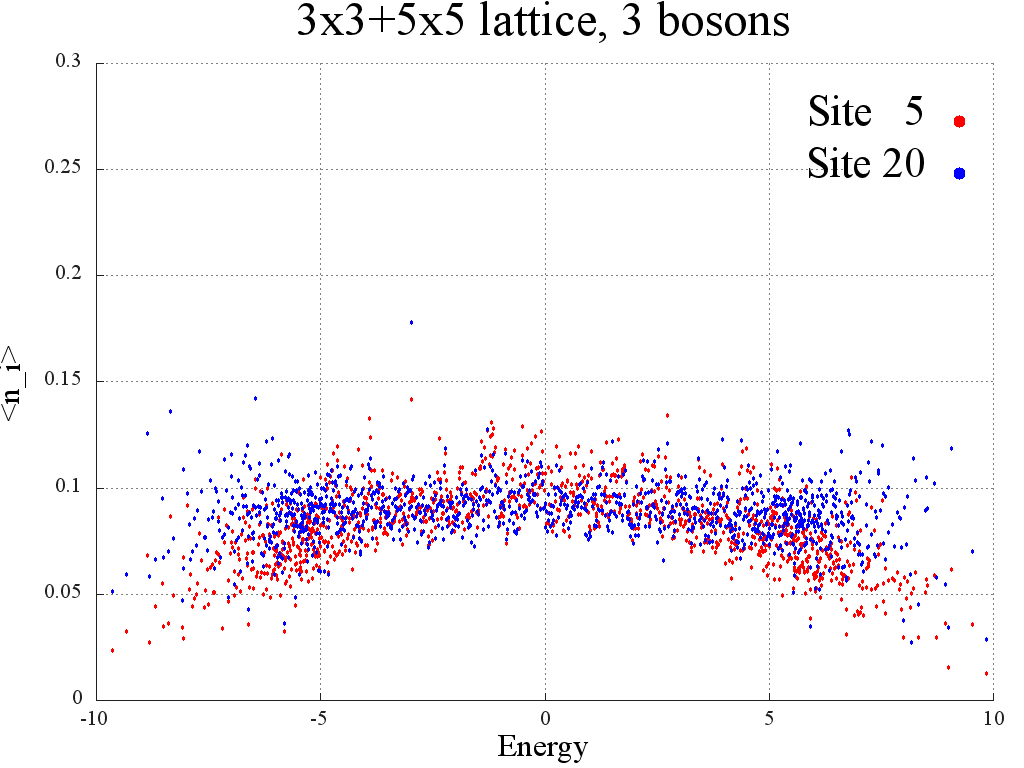}}
\subfloat[$N=53130, n=1240$   ]{\includegraphics[width=6.5cm]{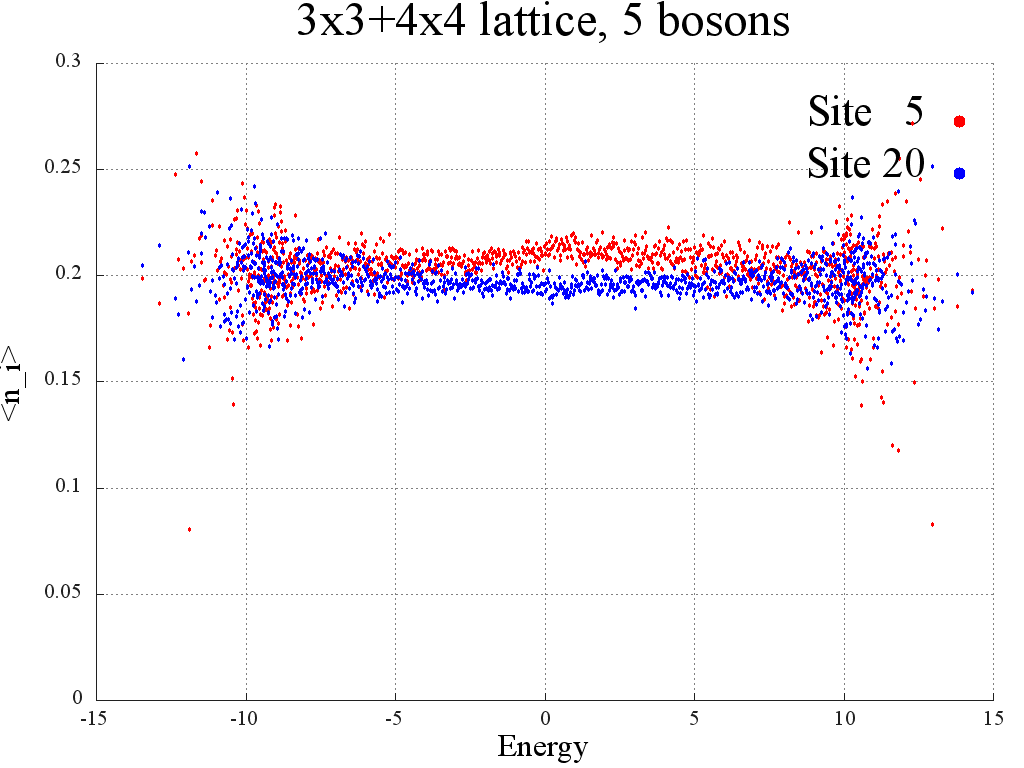}} \\
\subfloat[$N=1221759, n=1240$ ]{\includegraphics[width=6.5cm]{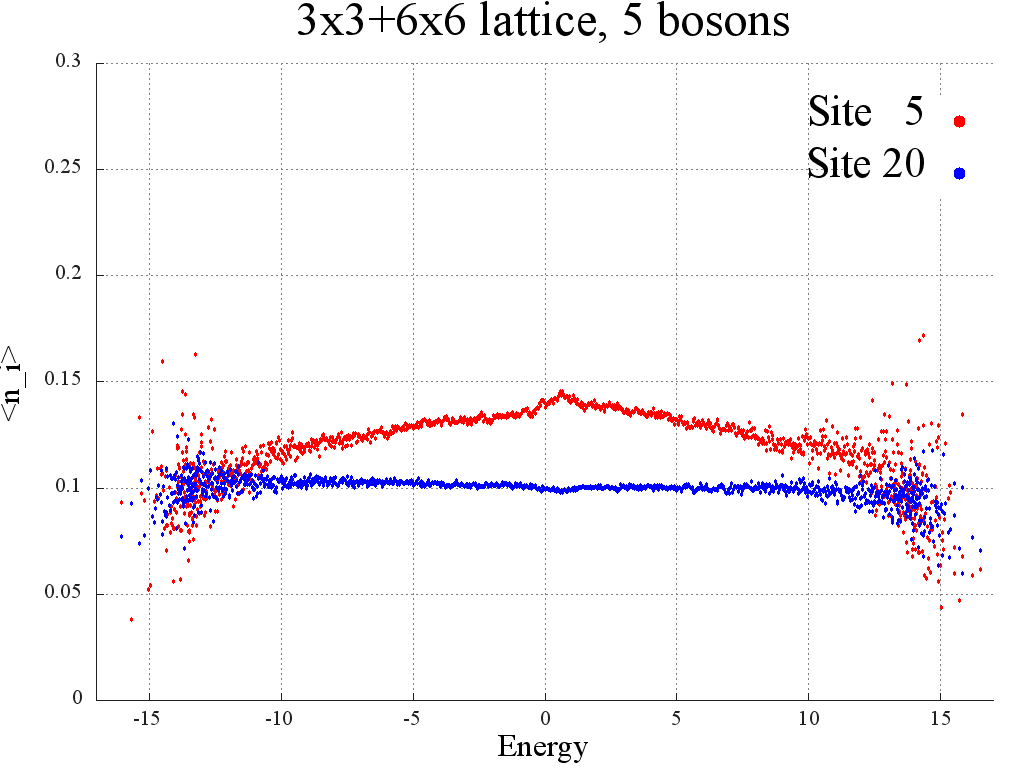}}
\subfloat[$N=15020334, n=1240$]{\includegraphics[width=6.5cm]{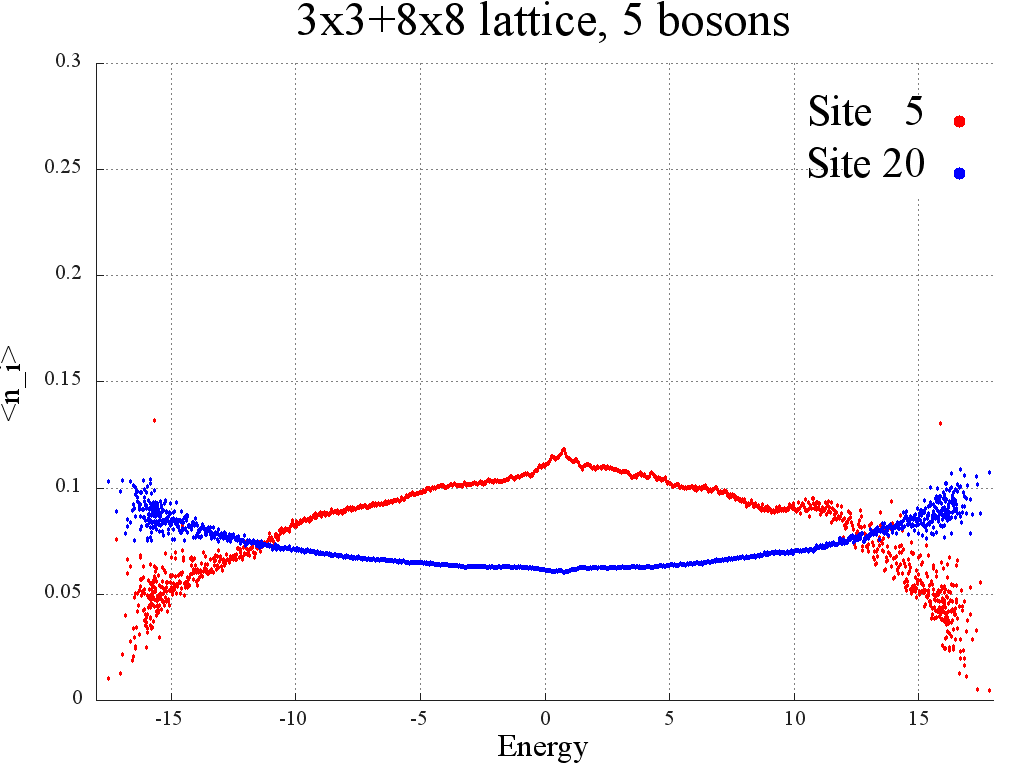}}
\caption{Mean value $\bra{\tilde{E}_\ell} n_i\ket{\tilde{E}_\ell}$ for sites $i=5$ and $i=20$ (see fig.\ref{Lattice}) as a function of the (Ritz) energy eigenvalue $\tilde{E}_\ell$ 
in a fixed dimension $n=1240$ Krylov subspace. The number $N$ indicates the dimension of the full Hilbert space. It is clearly seen that the function 
becomes smooth as $N$ becomes larger.}
\label{KrylovETH}
\end{figure}

 The operators we study first are the occupation numbers of the individual sites.
These are known to thermalize as shown in fig.\ref{Thermalization}. 
To check that the diagonal elements are
smooth functions of the energy we compute those functions for different lattice sizes and the
same size of the Krylov subspace. The results are shown
in fig.\ref{KrylovETH}. It is clear that the functions become smooth as we increase the size of the underlying Hilbert space. 

 Although we have shown only a few plots, the pattern is similar for other sites of the system and other initial states. Near the edges of the spectrum the occupation
number does not become smooth implying that at very low temperatures the finite system does not thermalize. In the intermediate region of the spectrum it is clear that
the mean value of the occupation number becomes a smooth function of the energy. 

\subsubsection{Single-particle occupation numbers}

As already mentioned, one motivation for considering $n_i$ as an operator suitable for ETH testing
is the special significance attached in statistical mechanics to local operators, as those characterizing subsystems
of a large isolated system. One may worry, however, that the smoothness of the average $n_i$ as a function of
energy is simply a consequence of translational invariance expected of the system in the thermodynamic limit.
 Indeed, consider a system that does not thermalize, for example free particles (fermions or bosons). In that case we can diagonalize the Hamiltonian 
for a single particle in a lattice of $N_s$ sites labeled by $i=1\ldots N_s$ (\eg\ the one in fig. \ref{Lattice}) obtaining eigenstates 
$k=1\ldots N_s$ with eigenfunctions $\psi_i^{(k)}$ and energy $\epsilon_k$. Define then single-particle creation operators $a^\dagger_k$ and the
corresponding occupation numbers:
\beq
 a^\dagger_k = \sum_i \psi_i^{(k)} a^\dagger_i, \ \ \ \ \ \  n_k = a^\dagger_k a_k  \, ,
\eeq 
where $a^\dagger_i$ creates a particle at site $i$. 
 The eigenstates of energy are 
\beq
 \ket{E_\nu} = \prod_k \ket{n_k} , \ \ \ \ E = \sum_k n_k \epsilon_k \, ,
\label{sing-part}
\eeq
where $n_k=0,1$ if the particles are fermions or any non-negative integer if they are bosons. 
  The expectation value of $n_i=a^\dagger_i a_i$ in this state is
\be
\bra{E_\nu} n_i \ket{E_\nu} = \sum_k n_k |\psi^{(k)}_i|^2 \, .
\label{psi2}
\ee
 In the case of low density studied here, with $N_p\ll N_s$ particles on the lattice, one expects that in the thermodynamic 
limit the average (\ref{psi2}) in a randomly chosen $\ket{E_\nu}$ approaches $N_p/N_s$, and the relative fluctuation of $n_i$ about it is of order $1/\sqrt{N_p}$. 
The reason is that, at low densities $N_p\ll N_s$, the typical state has $n_k=0,1$ even for bosons and there are $N_p$ terms in the sum in eq.(\ref{psi2}).
  Since this is the case for non-interacting particles, that are not expected to thermalize,
one may argue that the smoothness of $\bra{E_\nu} n_i \ket{E_\nu}$ is not a good measure of the ETH.\footnote{ 
Notice, however, that according to the ETH 
variation of $\bra{E_\nu} n_i \ket{E_\nu}$ from one $\nu$ to the next
should be suppressed {\em exponentially} in the number of particles, as opposed to a power law
\cite{Deutsch}.}

\begin{figure}
\centering
\subfloat[$N=5,984, n=1240$    ]{\includegraphics[width=6.5cm]{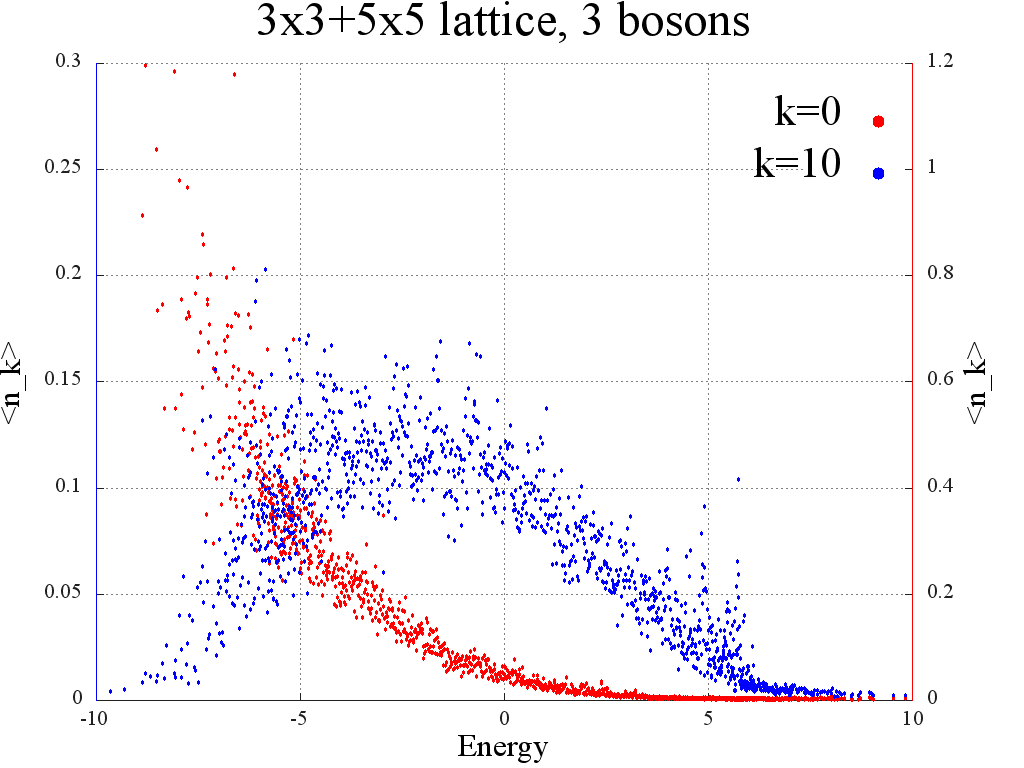}}
\subfloat[$N=53,130, n=1240$   ]{\includegraphics[width=6.5cm]{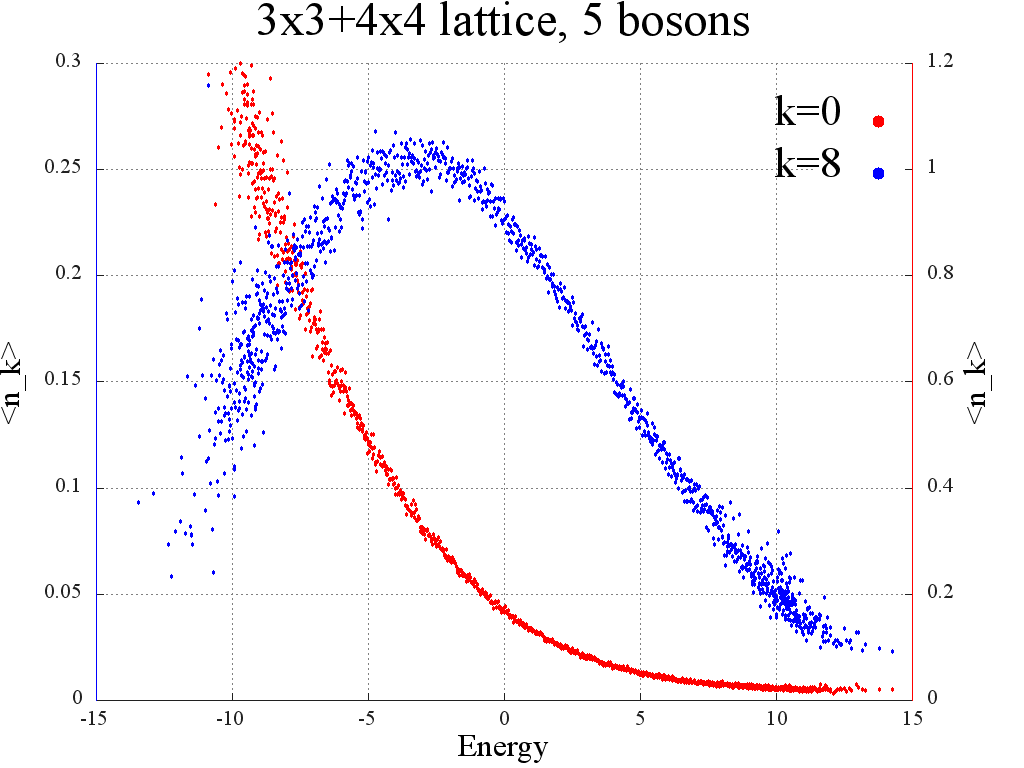}} \\
\subfloat[$N=1,221,759, n=1240$ ]{\includegraphics[width=6.5cm]{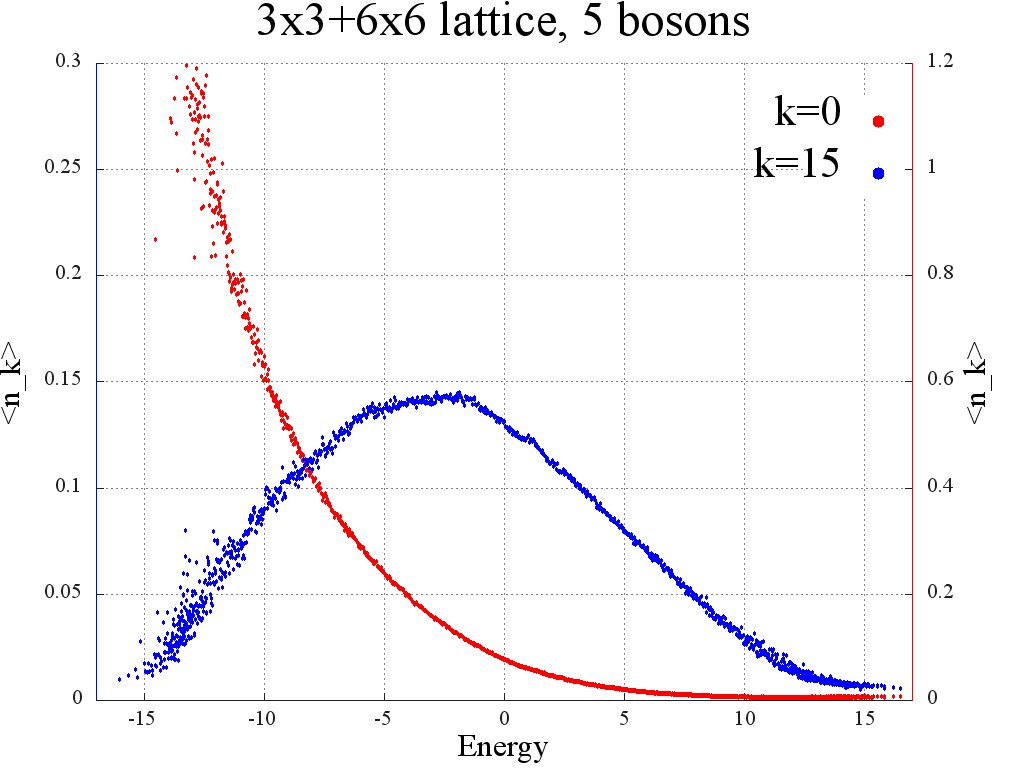}}
\subfloat[$N=15,020,334, n=1240$]{\includegraphics[width=6.5cm]{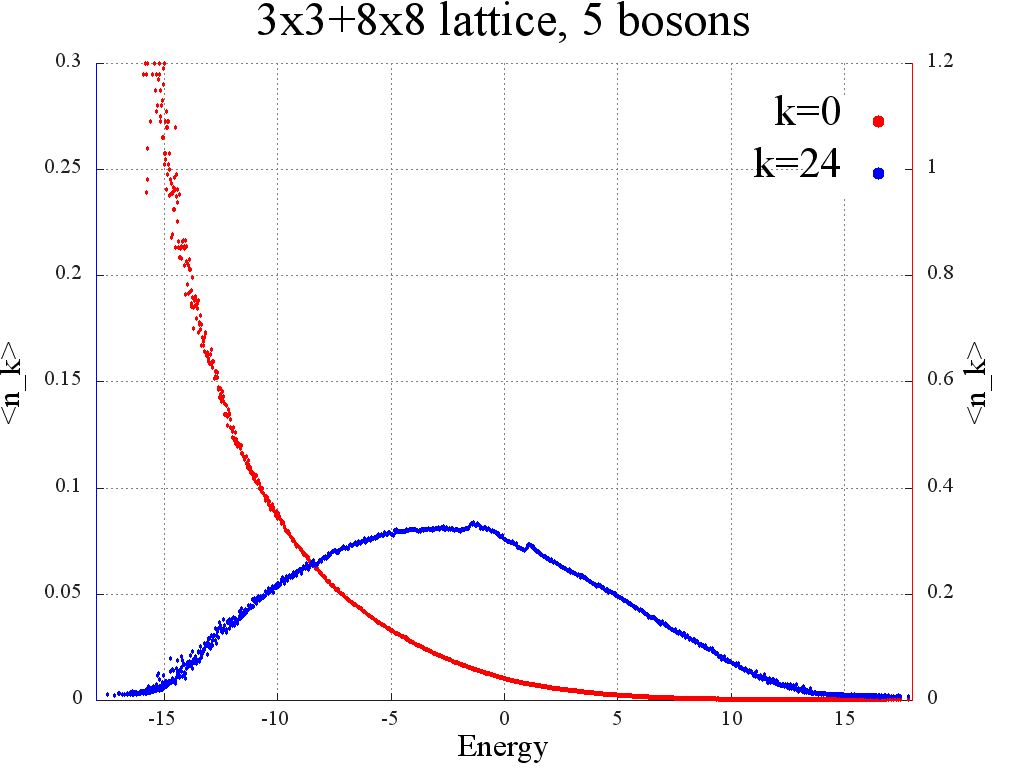}}
\caption{Mean values $\bra{\tilde{E}_\ell} n_k\ket{\tilde{E}_\ell}$ of the occupation numbers for two different
single particle states  as functions of the (Ritz) energy eigenvalue $\tilde{E}_\ell$ 
in a fixed dimension $n=1240$ Krylov subspace. The number $N$ indicates the dimension of the full Hilbert space. It is clearly seen that the function 
becomes smooth as $N$ becomes larger. The single particle energies are respectively
(a) $e_{k=0}=-3.49$, $e_{k=10}=-0.9$
(b) $e_{k=0}=-3.3$,  $e_{k=8 }=-0.89$
(c) $e_{k=0}=-3.61$, $e_{k=15}=-0.76$
(d) $e_{k=0}=-3.76$, $e_{k=24}=-0.7$.
}
\label{KrylovETHnk}
\end{figure}

 There are two ways to alleviate this concern. The first is to compute the average occupation numbers $\bra{E_\nu} n_k \ket{E_\nu}$ for 
the interacting case and see if they are smooth functions of the energy $E_\nu$. If so, that would distinguish the system from the non-interacting case,
where ETH is clearly not valid since $\bra{E_\nu} n_k \ket{E_\nu}$ takes only integer values and therefore cannot be a smooth function. Further, the $n_k$'s are interesting 
quantities in their own right, especially because in our case, at late stages of the evolution, the average density is low, so we may expect
our system to become a nearly ideal gas. In this case, as is well known, each single-particle mode can be considered 
as a separate subsystem, weakly coupled to the rest. As was the case for the $n_i$'s, since 
we do not know the exact eigenstates $\ket{E_\nu}$, we plot the expectation values of $n_k$
for the Ritz vectors $\ket{\tilde{E}_\ell}$, which are the relevant states for the thermalization of the initial state we considered. 
The result is shown in fig.\ref{KrylovETHnk} confirming once again that, in the interacting case, the ``Krylov ETH'' applies.

\subsubsection{Comparison with free fermions}

\begin{figure}
\centering
\subfloat[$N=5,984$, $5000$ states]{\includegraphics[width=6.5cm]{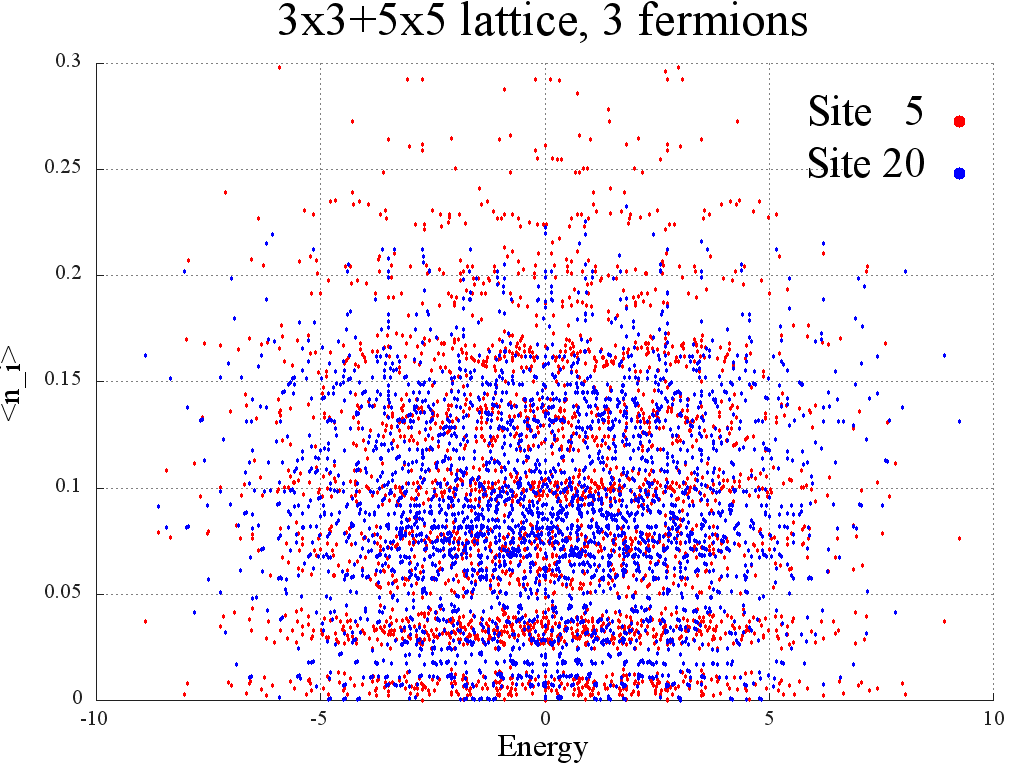}}
\subfloat[$N=53,130$, $5000$ states]{\includegraphics[width=6.5cm]{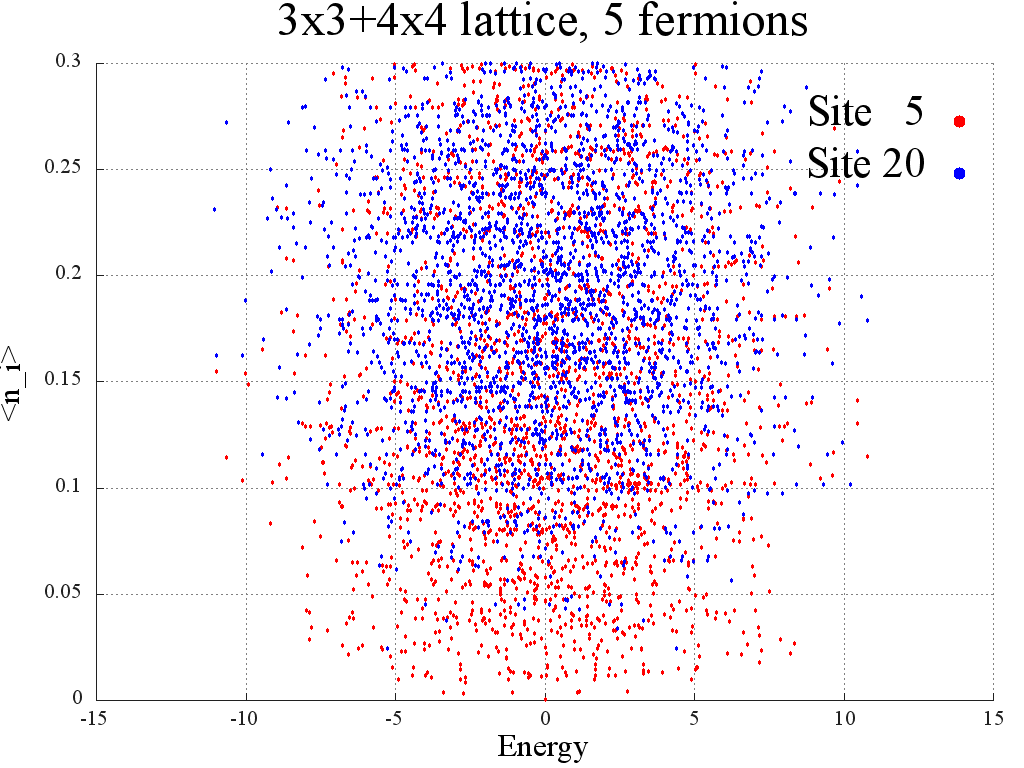}} \\
\subfloat[$N=1,221,759$, $5000$ states]{\includegraphics[width=6.5cm]{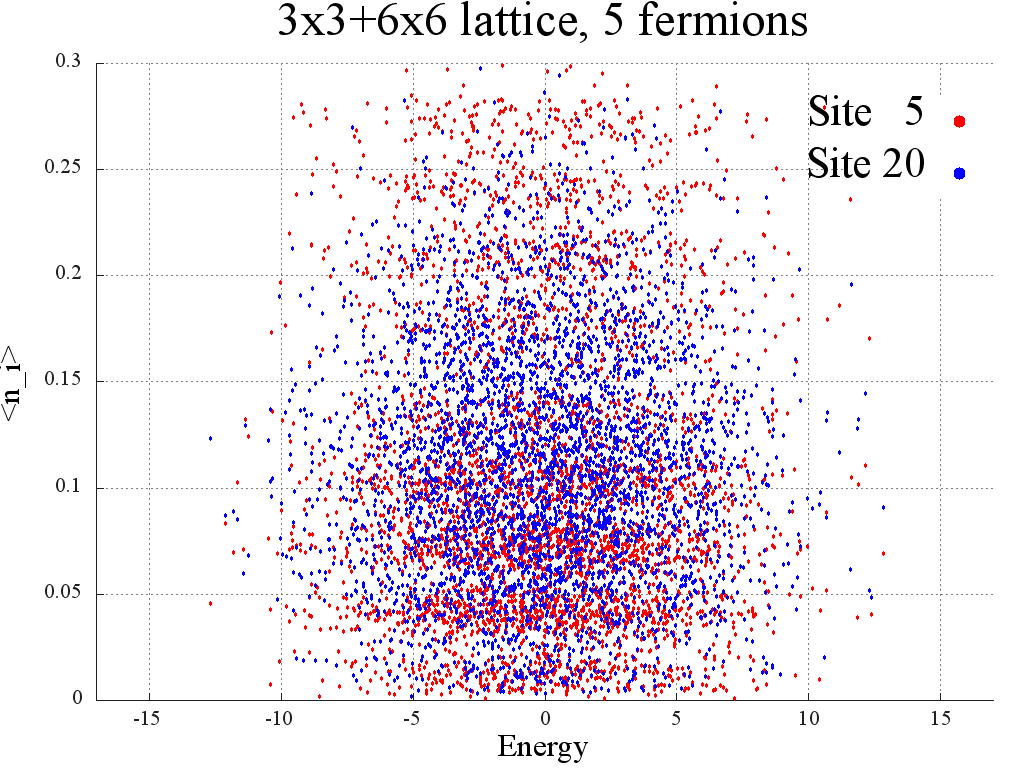}}
\subfloat[$N=15,020,334$, $5000$ states]{\includegraphics[width=6.5cm]{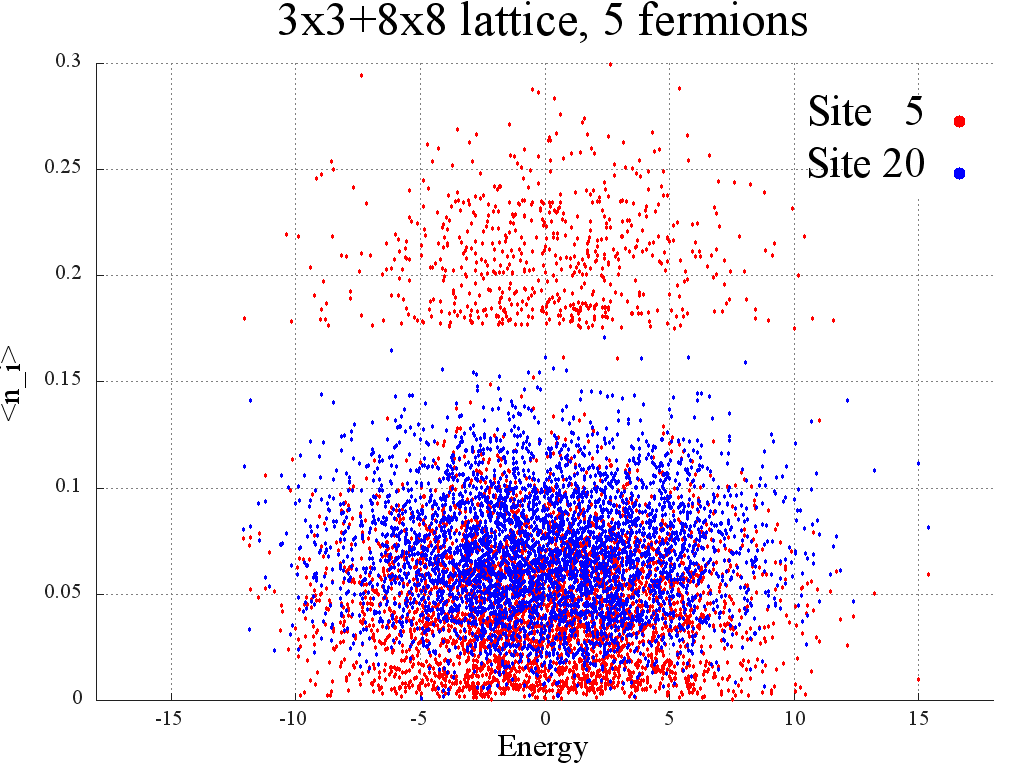}}
\caption{Mean value $\bra{E_\nu} n_i\ket{E_\nu}$ for sites $i=5$ and $i=20$ (see fig.\ref{Lattice}) as a function of the energy eigenvalue $E_{\nu}$ 
for a random sample of exact eigenstates for the case of free fermions. Since the system is exactly solvable we could plot all eigenstates but 5000 points
is already enough to show that there is no smooth function relating the mean occupation number and the energy.}
\label{NiFermions}
\end{figure}

 The second way to differentiate the interacting case from a non-interacting one 
is simply to plot, as a function of the
eigenstate energy, the site occupation numbers for 
a system of free fermions\footnote{We use free fermions since their site occupation numbers are $n_i=0,1$ as for hard bosons.} and compare them to the results for interacting bosons.
Since, in this case, we can find the exact eigenstates, instead of using the Krylov subspace, a random sample of 5000 eigenstates was used to produce the plots displayed in fig.\ref{NiFermions}.
 The occupation number as a function of energy does not fall at all on a smooth curve. Comparing fig.\ref{NiFermions} and fig.\ref{KrylovETH} one can see the predictive power of the ETH. 
We should, however, point out some caveats. In the free fermion figure we used exact eigenstates, whereas in fig.\ref{KrylovETH} we used the Ritz states (as we are not
able to compute exact ones for large systems). Finally, what we see here is that similar, local operators behave differently in the interacting and non-interacting
cases. There may be other, presumably non-local, operators that obey the ETH property even in the non-interacting case but would not be ordinarily of interest in applications 
of statistical mechanics. That is, the ETH property is a property of the system as defined by the Hamiltonian and a set of operators that we want to measure and not a property of the 
Hamiltonian by itself.  

\subsubsection{Off-diagonal matrix elements}

Return to the interacting case and consider an operator $\hat{A}$ that thermalizes, for
example, the site occupation $n_i$.
According to the ETH, the off-diagonal elements in the basis of the exact energy eigenstates
are small, an estimate being \cite{Srednicki}
\beq
 \bra{E_\nu} \hat{A} \ket{E_\nu'} \sim \frac{\bar{A}}{\sqrt{N}}, \ \ \ \ \nu\neq\nu' \, ,
\label{off-diag-exact}
\eeq
where $\bar{A}$ is the magnitude of the diagonal elements (used here to fix the scale), 
and $N$ is the dimension of 
the full Hilbert space. Numerically, we have to consider Ritz vectors $\ket{\tilde{E}_\ell}$ 
instead of the exact eigenstates 
$\ket{E_\nu}$. What can we expect for the off-diagonal elements of those?

First, recall that the reason for postulating that the off-diagonal
elements (\ref{off-diag-exact}) are small is that, beyond the thermalization time, 
the expectation value (\ref{a2})
has to become time-independent. When we follow the
time evolution using the Krylov subspace, we obtain the counterpart to (\ref{a4}) 
in which the exact eigenstates are replaced with the Ritz vectors, and $E_\nu$ with $\tE_\ell$.
However, if the Krylov subspace has $n$ states, we can use it to follow the evolution
only until times of order $t\sim n  t_0$, where $t_0$ is some constant time that fixes 
the scale. For that reason, the exponential terms $\exp[-i(\tE_\ell - \tE_{\ell'})t]$
with energy differences much smaller than $1/(n t_0)$ should be considered
constant, and the corresponding off-diagonal matrix elements do not have to be small. 

In fact, we must expect a total $O(1/n)$ contribution to 
$\bra{\psi(t)} \hat{A} \ket{\psi(t)}$ from off-diagonal
$\langle \tE_{\ell'}| \hat{A} | \tE_\ell\rangle$ with $\tE_{\ell'}$ close to $\tE_\ell$.
This is the accuracy to which, according to the estimate (\ref{estRitz}), the
diagonal element $\tilde{A}(\tE_\ell)$ approximates the exact one, $A(E_\nu)$.
Since, as we have seen, the time evolution can be followed using the Krylov subspace methods 
much more accurately than that, we have to conclude that the error in the diagonal elements 
must be compensated by the contribution from the off-diagonal ones.

An estimate for $\langle \tE_{\ell'}| \hat{A} | \tE_\ell\rangle$ with 
$\tE_{\ell'} \approx \tE_\ell$ can be obtained in the
same way as we have obtained (\ref{estRitz}). Namely, neglect the $O(1/\sqrt{N})$
quantities (\ref{off-diag-exact}) altogether, so that
\be
\bra{\tE_{\ell'}} \hat{A} \ket{\tE_\ell} \simeq \sum_\nu c^*_{\ell' \nu} c_{\ell \nu}
A(E_\nu) \, ,
\label{off-diag-Ritz}
\ee
and expand $A(E_\nu)$ in Taylor series near $\tE_\ell$. As before,
$c_{\ell \nu} = \langle E_\nu | \tE_\ell\rangle$. In addition, we will need the following
property of the Ritz states (to be derived in the next section):
\be
H \ket{\tE_\ell} = \tE_\ell \ket{\tE_\ell} + \ket{\xi_\ell} \, ,
\label{xi}
\ee
where the ``residual'' $\ket{\xi_\ell}$ is orthogonal to the entire Krylov subspace, and
the residuals for different $\ell = 0,\ldots,n-1$ are parallel, i.e., differ
only by overall factors. The result is
\be
|\bra{\tE_{\ell'}} \hat{A} \ket{\tE_\ell}| = 
\half |A''(\tE_\ell)| \norm{\ket{\xi_\ell}} \norm{\ket{\xi_{\ell'}}} + \ldots \, .
\label{est-off-diag}
\ee
One readily sees that $\norm{\ket{\xi_\ell}} = \Delta E_\ell$, 
the standard deviation of energy in the
Ritz state $\ket{\tE_\ell}$. This will be argued shortly to be $O(1/\sqrt{n})$; hence,
(\ref{est-off-diag}) is $O(1/n)$.

\begin{figure}
\centering
\subfloat[$N=5,984$]{\includegraphics[width=6.5cm]{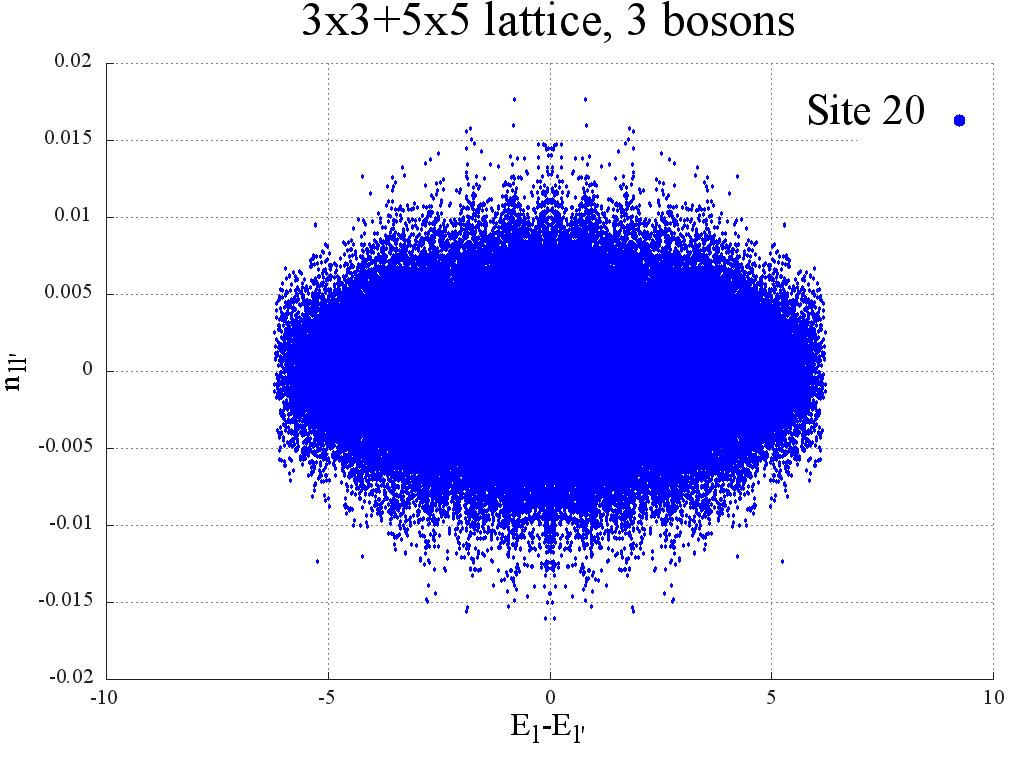}}
\subfloat[$N=53,130$]{\includegraphics[width=6.5cm]{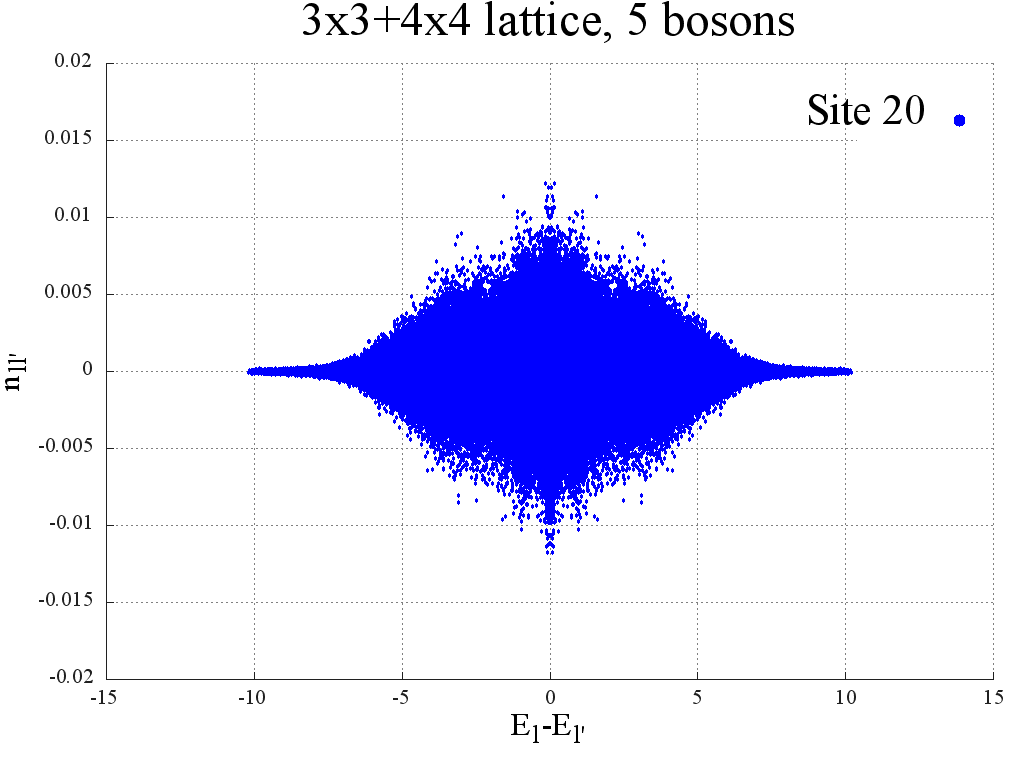}} \\
\subfloat[$N=1,221,759$]{\includegraphics[width=6.5cm]{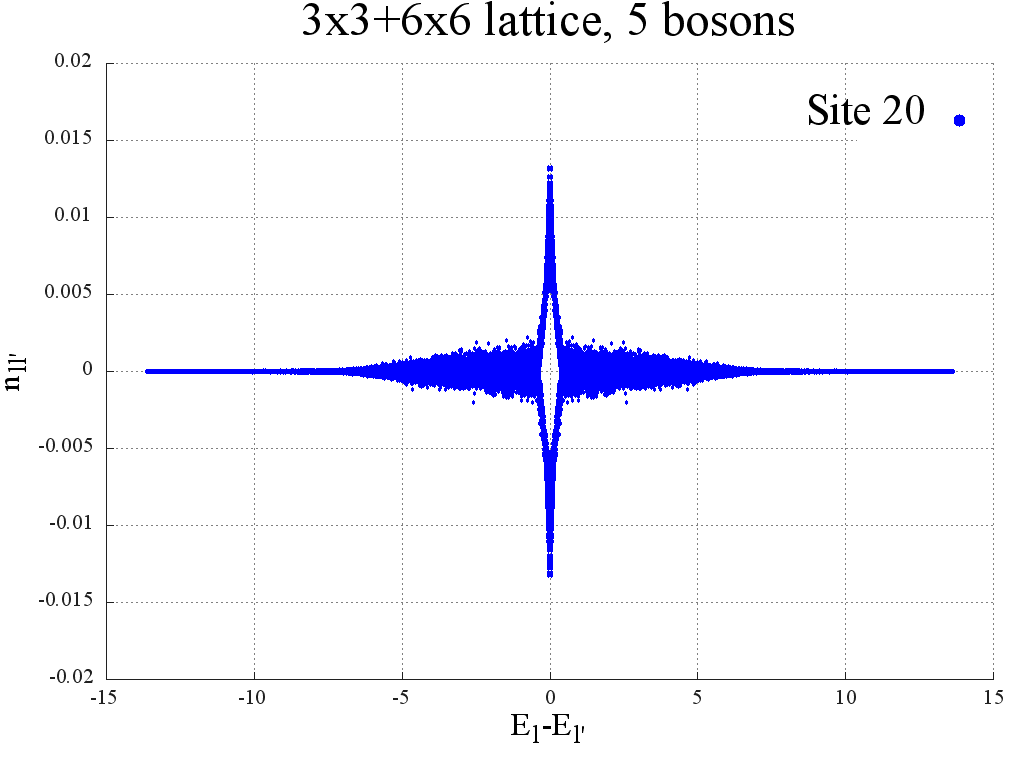}}
\subfloat[$N=15,020,334$]{\includegraphics[width=6.5cm]{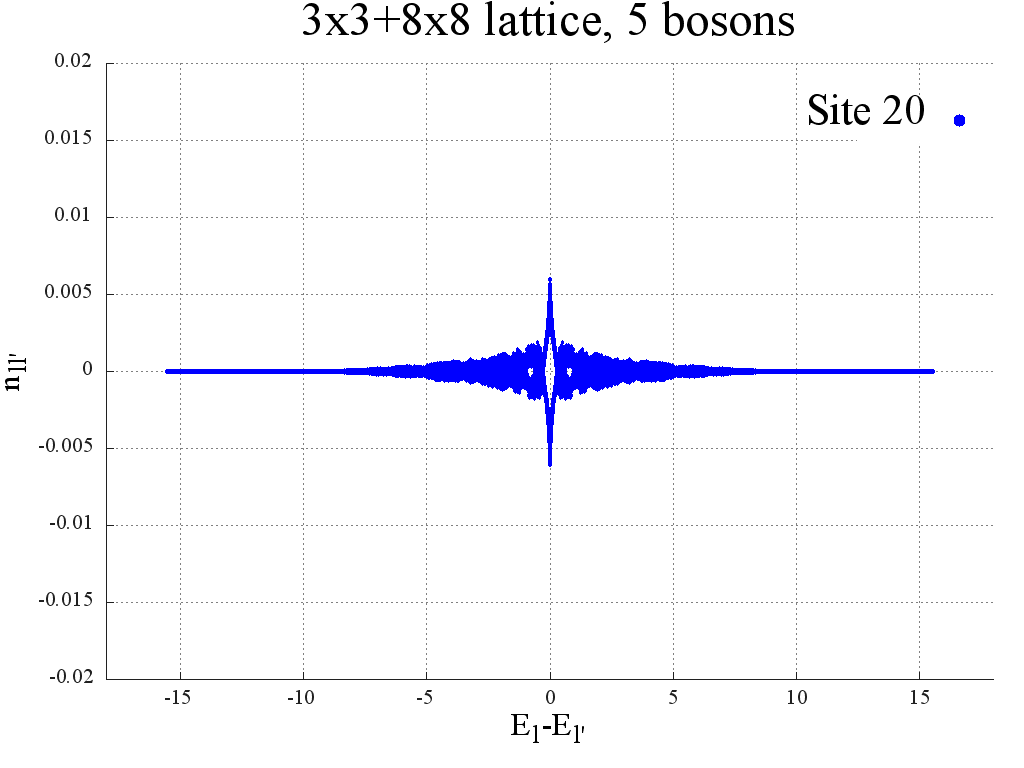}}
\caption{Off-diagonal elements $\bra{\tilde{E}_\ell} n_i\ket{\tilde{E}_{\ell'}}$ for site $i=20$ (see fig.\ref{Lattice}) as a function of the energy difference $E_{\ell}-E_{\ell'}$ 
for all Ritz vectors $\ket{\tilde{E}_\ell}$ in a Krylov subspace of order $N_K=1240$.}
\label{off-diag}
\end{figure}

Numerically computed values of the
off-diagonal elements $\bra{\tilde{E}_\ell} n_i \ket{\tilde{E}_{\ell'}}$ for a given site 
are plotted, as functions of the energy difference $(\tilde{E}_\ell-\tilde{E}_{\ell'})$, 
in figure \ref{off-diag}. It is clear that, as we increase the size of the total Hilbert space
they become smaller, except in a region around the diagonal (center of the plot) where, as we discussed before, they do not need to be small.  

%\begin{figure}
%\centering
%\subfloat[$N=5,984$]{\includegraphics[width=6.5cm]{plotNijNs34Nb3Nc2NK1240.png}}
%\subfloat[$N=53,130$]{\includegraphics[width=6.5cm]{plotNijNs25Nb5Nc2NK1240.png}} \\
%\subfloat[$N=1,221,759$]{\includegraphics[width=6.5cm]{plotNijNs45Nb5Nc2NK1240.png}}
%\subfloat[$N=15,020,334$]{\includegraphics[width=6.5cm]{plotNijNs73Nb5Nc2NK1240.png}}
%\caption{Off-diagonal elements $\bra{\tilde{E}_\ell} n_i\ket{\tilde{E}_{\ell'}}$ for site $i=20$ (see fig.\ref{Lattice}) as a function of the energy difference $E_{\ell}-E_{\ell'}$ 
%for all Ritz vectors $\ket{\tilde{E}_\ell}$ in a Krylov subspace of order $N_K=1240$.}
%\label{off-diag}
%\end{figure}

\subsubsection{Bose-Einstein distribution}
\begin{figure}
\centering
\includegraphics[width=10cm]{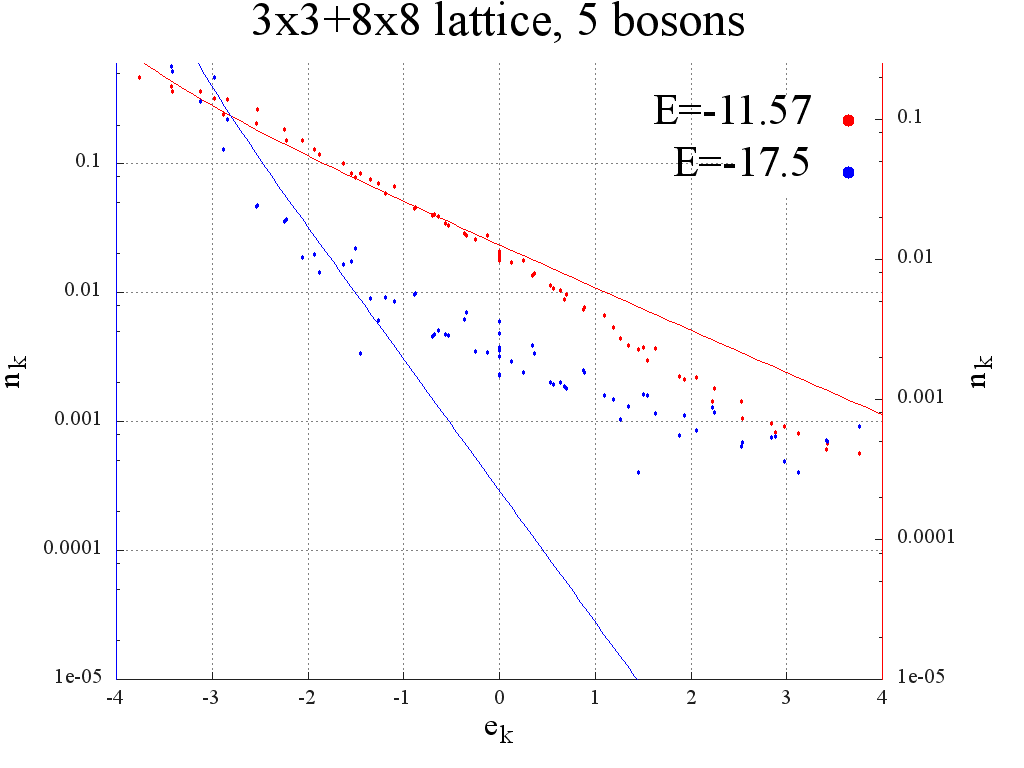}
\caption{Occupation numbers $n_k$ of different single-particle eigenstates as functions of 
the single-particle energies $e_k$ for two different Ritz states of energies $\tilde{E}=-17.5$ and $E=-11.57$. The state with the lower
energy is the ground state. The approximately straight lines are the corresponding 
Bose-Einstein distributions with inverse temperature $\beta$ and chemical potential $\mu$ fitted to the total energy and number of bosons ($\beta=2.16, \beta\mu=-8.41$ for $\tilde{E}=-17.5$ and $\beta=0.754, \beta\mu=-3.77$ for $\tilde{E}=-11.57$). The ground state is clearly not thermal but the excited state is.}
\label{RitzETH}
\end{figure}

Finally, we can perform one more test of the ETH. This hypothesis implies that even 
if the whole system is in 
an energy eigenstate the behavior of local, and also few body, operators is thermal. 
For a model of qubits with a random nearest-neighbor interaction, this has been
described \cite{QCapp} as the (sufficiently strong) 
interaction playing the role of a thermal bath for
the individual qubits. Here, we can test if a particular few-body operator behaves 
thermally when the entire system is in a single Ritz state.
In fig.~\ref{RitzETH} we plot the occupation number of single-particle eigenstates 
as a function of 
the single-particle energy for the system with $N_s=73$ sites and $N_b=5$ bosons. We expect that, in thermal equilibrium,
this system is dilute enough to be close to an ideal gas. 
In that case, the occupation numbers should be given by the Bose-Einstein distribution
\beq
 n(\epsilon_k) = \frac{1}{e^{\beta\epsilon_k-\beta\mu}-1} \, ,
\eeq
where the inverse temperature $\beta$ and the chemical potential $\mu$ are fixed by the total energy and particle number.
Note that there are no parameters left to fit the distribution. One of the plots in fig.~\ref{RitzETH}
corresponds to the ground state,
which the Lanczos method finds essentially exactly, and the other to an excited state in the lower half of the spectrum
(we only plot the results for one but most behave similarly)\footnote{Except that states in the upper half of the spectrum have
negative temperatures}. We see that the ground state is not thermal. This can be expected: 
the ends of the spectrum do 
not obey the ETH. On the other hand, the excited state behaves thermally, except for large
single-particle energies. We attribute the discrepancy at large $\epsilon_k$ to the
total number of bosons being small. 
 
 To summarize this section, we conclude that the ETH has passed all the numerical tests performed in this paper.

\section{Lanczos matrix}

We have seen that to describe the approach of a system, initially in a pure state
$|\psi_0\rangle$, to equilibrium, it is sufficient to 
diagonalize the Hamiltonian projected onto a sufficiently large Krylov subspace (KS):
\be
\tH = \PK{n} H \PK{n} \, ,
\label{tH}
\ee
where $H$ is the original Hamiltonian and $\PK{n}$ is the projector onto the KS 
spanned by the states\footnote{In this section, $i$ and $j$ label the basis states in the
Krylov subspace, rather than the sites of the physical lattice.}
\be
|\psi_i\rangle = H^i |\psi_0 \rangle \, , \hspace{3em} i = 0,\ldots, n-1 \, .
\ee
For a macroscopic
system, a``sufficiently 
large'' KS is still a tiny fraction of the entire Hilbert space. 

Suppose we have an orthonormal basis in the KS, formed by states $|q_i\rangle $, $i=0,\ldots,n-1$.
These are some linear combinations of the states $|\psi_i\rangle$ above. Each
$|q_i\rangle$ belongs to the full Hilbert space, of the large dimension $N \gg n$.
Using these states,
the projector in (\ref{tH}) can be written as
\be
\PK{n} = \sum_{i=0}^{n-1} |q_i\rangle \langle q_i | \, .
\ee
In the basis of $\ket{q_i}$, the projected Hamiltonian is represented by the matrix 
\be
\tH_{ij} = \langle q_i | \tH | q_j \rangle \, .
\label{tHmat}
\ee
Let the eigenvectors of this matrix be some $r_\ell$:
\be
\tH r_\ell = \tE_\ell r_\ell \, ,
\label{evecs}
\ee
$\ell=0,\ldots n-1$.
These are ``short'' vectors, of dimension $n$. They can be assembled into ``long'' ones, of 
dimension $N$, as follows:
\be
|\tE_\ell \rangle = \sum_{i=0}^{n-1} r_{\ell i} |q_i\rangle \, ,
\label{ritz}
\ee
where $r_{\ell i}$ is the $i$th component of $r_\ell$.
These are the Ritz vectors already discussed in the preceding sections.
Clearly, if $n=N$, the Ritz vectors are
eigenstates of the original Hamiltonian. Our goal in this section is to see in what sense 
they can be thought to represent the true eigenstates when $n \ll N$.

In principle, there are many different ways to choose the orthonormal basis $|q_i\rangle $. Here,
we adopt the Lanczos method \cite{Lanczos}, in which $|q_i\rangle $ are such that the matrix (\ref{tHmat}) is 
tridiagonal. 
%This method is 
%fairly economical in cases, such as ours, when the original $H$, represented as an $N\times N$ matrix,
%that the resulting tridiagonal matrix has certain properties, which are
%useful for a more precise understanding of how the Ritz vectors $\ket{\tE_l}$ 
%are related to the true eigenstates
%$\ket{E_\nu}$. 

The Lanczos method can be defined as a sequence of iterations, operating with the original
Hamiltonian $H$, such that upon the $(n-1)$st iteration we would have constructed the projected 
$\tH$, together with the basis $|q_i\rangle $ in which it is tridiagonal. Suppose 
we wish to find
a unitary $N\times N$ matrix $Q$ that reduces the full $H$ to a tridiagonal $T$:
\be
T = Q^\dagger H Q \, .
\label{T}
\ee
Our notation for the elements of $T$ is as follows:
\be
T = \left( \begin{array}{cccc} 
\alpha_0 & \beta_1 & 0 & \ldots \\
\beta_1^* & \alpha_1 & \beta_2 & 0 \\
0 & \beta_2^* & \alpha_2 & \beta_3 \\
\ldots
\end{array} \right)
\label{Tmat}
\ee
Denote the $i$th column of $Q$ by $|q_i\rangle$. 
Rewrite (\ref{T}) as
\be
H Q = Q T \, ,
\label{QT}
\ee
and pick the $i$th column of this. The result is
\be
\beta_{i+1}^* |q_{i+1} \rangle = 
(H - \alpha_i) |q_i\rangle - \beta_i |q_{i-1}\rangle \, ,
\label{rec}
\ee
where by definition $\beta_0 \equiv 0$. Orthogonality among $|q_i\rangle$ implies
\ba
\alpha_i & = & \langle q_i | H |q_i\rangle \, , \label{alpha} \\
\beta^*_{i+1} & = &  \langle q_{i+1} | H |q_i\rangle \label{beta} \, .
\ea
Choose $|q_0\rangle = |\psi_0\rangle$, the initial state of the system. Then, at the $(i+1)$st step,
$\beta^*_{i+1}$ and  $|q_{i+1} \rangle$ are determined by normalizing the right-hand side of the Lanczos recursion (\ref{rec}). 
Thus, after $(n-1)$ steps our vectors $|q_i\rangle$ span precisely the $n$-dimensional Krylov subspace. 

 Although at each step only the two previous vectors $\ket{q_i}$ and $\ket{q_{i-1}}$ are used, the Lanczos procedure in exact arithmetic guarantees that the new vector $\ket{q_{i+1}}$ is orthogonal to all the previous ones. 
 When done in machine arithmetic, however, 
orthogonality among the basis vectors is lost after some steps \cite{Lanczos}. For that reason every certain number of steps the new vector $\ket{q_{i+1}}$ is explicitly made
orthogonal to all the previous ones. This is the procedure used in this paper. 

 Let us now estimate how well the Ritz vectors (\ref{ritz}) represent the eigenstates 
of the original
problem. In the basis generated by the Lanczos method, the projection (\ref{tH}) 
amounts simply to
retaining only the $n\times n$ upper left corner of the matrix (\ref{Tmat}).
Then, the $i$th row of the 
eigenvalue equation (\ref{evecs}) is
\be
\beta^*_i r_{\ell,i-1} + \alpha_i r_{\ell i} + \beta_{i+1} r_{\ell,i+1} = \tE_\ell r_{\ell i}  \, ,
\ee
for $i = 0,\ldots, n-2$, and 
\be
\beta^*_{n-1} r_{\ell,n -2} + \alpha_{n-1} r_{\ell,n-1}  = \tE_\ell r_{\ell, n-1}  \, ,
\ee
for $i = n-1$. Multiplying these by $|q_i\rangle$, summing over all $i$, and using the 
recursion relation (\ref{rec}), we obtain
\be
H |\tE_\ell\rangle = \tE_\ell |\tE_\ell\rangle + \beta^*_n r_{\ell,n-1}  |q_n\rangle \, ,
\ee
for the Ritz vector (\ref{ritz}). Thus, the variance of energy in the Ritz state,
\be
(\Delta E_\ell)^2 \equiv 
\langle \tE_\ell | (H - \tE_\ell)^2 |\tE_\ell\rangle = |\beta_n|^2 |r_{\ell,n-1}|^2 \; ,
\label{var}
\ee
is determined by the matrix element $\beta_n$ and the last component of the eigenstate
of the reduced problem (\ref{evecs}).

If the variance (\ref{var}) is close to zero, it means that the corresponding Ritz vector
is close to a true eigenstate of $H$. We refer to such Ritz vectors as having converged
(to some specified precision). In particular, we have observed that the values
of $\beta_n$ are not particularly small, at least not until the size $n$ of the Krylov
subspace approaches the total Hilbert space dimension $N$. In other words, convergence
of Ritz vectors
is due to smallness of $r_{\ell,n-1}$, not of $\beta_n$. This leads us to the following analogy.
Consider the tridiagonal matrix of the reduced Hamiltonian $\tH$ (the $n\times n$
upper left corner of the matrix $T$) as a Hamiltonian of a fictitious particle hopping
along a 1-dimensional chain with $n$ sites, labeled by $i=0,\ldots,n-1$. Then,
$\beta_i$ correspond to
the hopping amplitudes, $\alpha_i$ to the on-site potential, and $r_{\ell i}$ to the
wave-function of the particle in the eigenstate number $\ell$. The Ritz vectors
that have already
converged correspond to $r_{\ell i}$ that are localized, i.e., decay rapidly towards
the right end of the chain, 
and those that are still far from convergence to $r_{\ell i}$ that are extended
over the entire chain.

\begin{figure}
\centering
\subfloat[$N=15020334, n\le 960$    ]{\includegraphics[width=6.5cm]{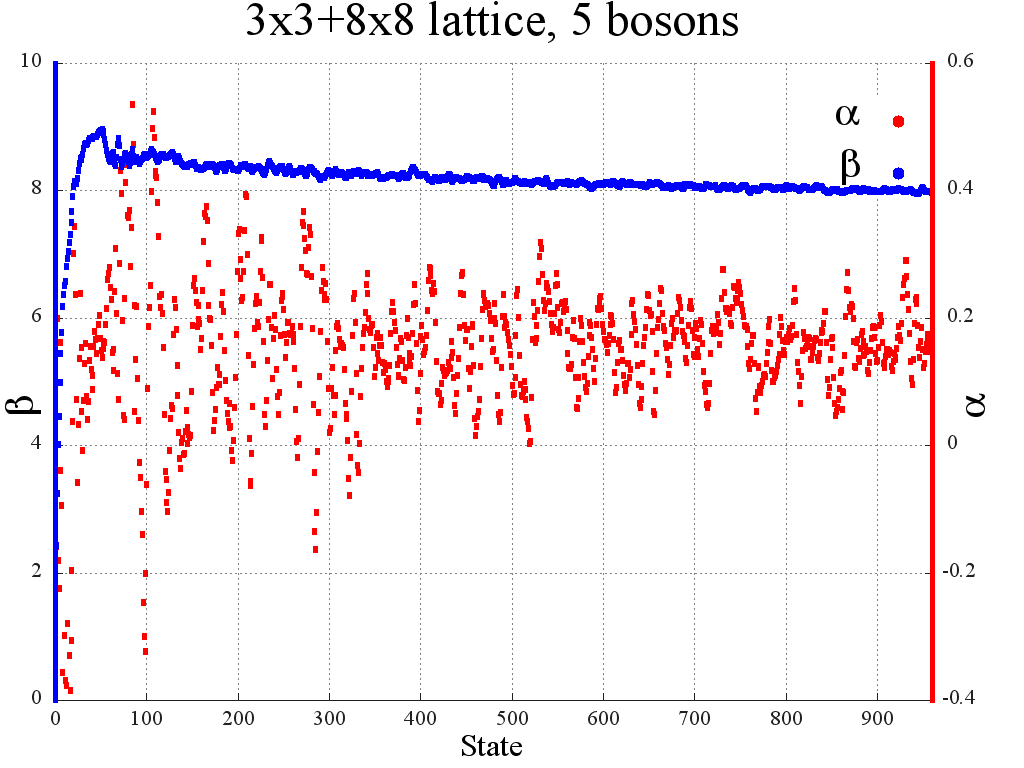}}
\subfloat[$N=10000, n\le N=10000$   ]{\includegraphics[width=6.5cm]{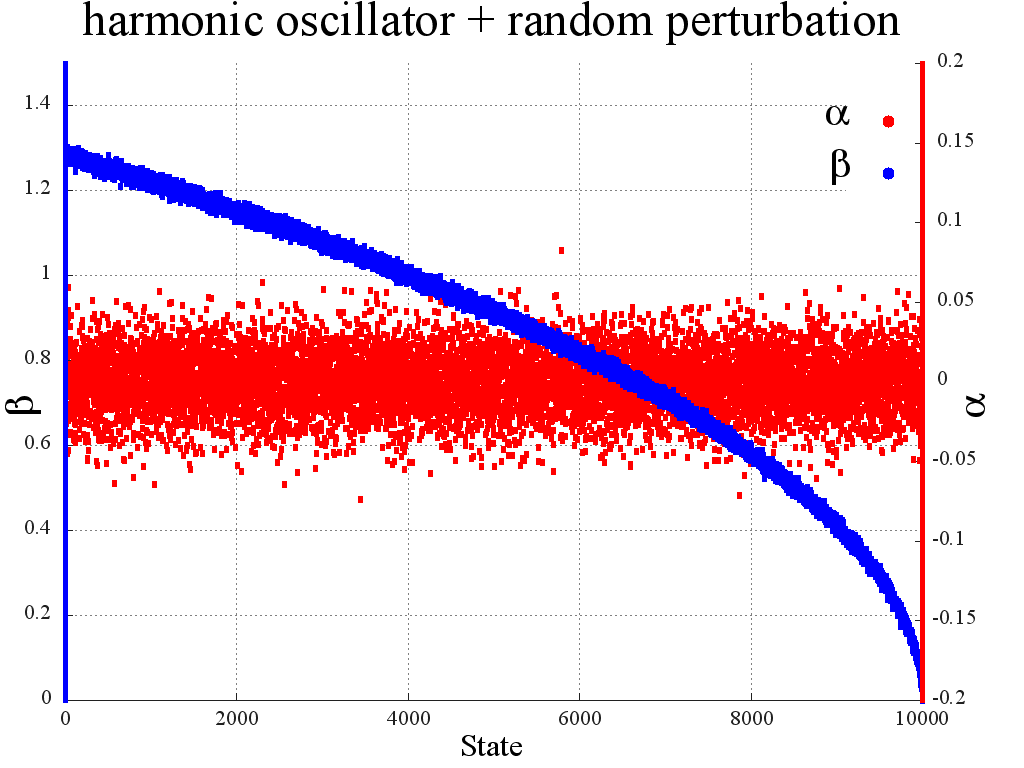}} 
\caption{Diagonal $\alpha_n$ and off-diagonal $\beta_n$ elements of the Lanczos matrix as function of the size of the Krylov subspace $n$ for (a) the 
lattice gas and (b) the harmonic oscillator (equidistant spectrum $-1\le E_i\le 1$) with a random perturbation ($h_{ij}\in[-0.02,0.02]$). Notice the 
different scales for $\alpha$ and $\beta$. Denoting by $N$ the size of the full Hilbert space, we see that for  $n/N\ll 1$, $\beta_n$ is approximately 
constant and $\alpha_n$ is randomly distributed. In the case of the lattice gas, the dispersion in $\alpha$ is larger than typical due to the particular 
initial state. In case (b), the initial state was chosen randomly. Note that $\beta_{n\rightarrow N} \rightarrow 0$, a behavior that we also verify in 
the lattice gas when it can be fully diagonalized.} 
\label{AlphaBeta}
\end{figure}

There is an analogy here with Anderson localization of electron in a disorder potential.
Indeed, variation of $\alpha_i$ and $\beta_i$ with $i$ means that there is both
site and bond disorder. We can roughly estimate the magnitude of variation in $\alpha_i$
as follows. Let the expansion of the Lanczos vector $|q_i\rangle$ in the eigenstates
$|E_\nu\rangle$ of the full Hamiltonian be
\be
|q_i\rangle = \sum_{\nu=0}^{N-1} c_{\nu i} |E_\nu \rangle \, .
\ee
Using this in (\ref{alpha}), we obtain
\be
\alpha_i = \sum_{\nu=0}^{N-1} |c_{\nu i}|^2 E_\nu \, .
\label{alpha2}
\ee
The values $\alpha_i$ for the first few $i$ depend on the initial state
$|q_0\rangle = |\psi_0\rangle$ and may exhibit some special structure
in the coefficients $c_{\nu i}$. We expect, however, that repeated application of the 
Hamiltonian during the Lanczos recursion rapidly spreads $c_{\nu i}$ over the entire
spectrum, essentially in a random manner.\footnote{Eqn.(\ref{rec}) shows that the
amplitudes $c_{\nu i}$ with $\nu$ near the edges of the spectrum, where
$H-\alpha_i$ is the largest, get amplified during the recursion. Note that these 
$c_{\nu i}$ are generically non-vanishing, except for a few 
$\nu$, for which the recursion has already converged, and the corresponding eigenstates $\ket{E_\nu}$
are linear combinations of only a finite number of states $\ket{q_i}$.
}
If, for a given $i$, $|c_{\nu i}|^2$ are random numbers
distributed uniformly between 0 and 1, the r.m.s. fluctuation of $\alpha$ is of order
\be
\alpha' \sim \frac{W}{\sqrt{N}} \, ,
\label{flucalpha}
\ee
where $W= \half (E_{\max} - E_{\min})$ is half the total
bandwidth. The average value of $\alpha$ is
close to zero, as the spectrum in our case is nearly symmetric about $E=0$ (the average is,
in any case, immaterial, as it only shifts the potential by a constant, without affecting
the localization properties). For $\beta_i$, we estimate the 
average as $\bbeta \sim W$ on dimensional grounds, and the fluctuation as
$\beta' \sim W / \sqrt{N}$, similarly to (\ref{flucalpha}). 
These estimates are well born out numerically, see
Fig.~\ref{AlphaBeta}. In the figure, we have also included the results of the Lanczos recursion 
for a harmonic oscillator with 
a random perturbation Hamiltonian and a random initial state, to 
show that the behavior discussed here is rather generic.

The conclusion we draw from these estimates is that a useful starting point for estimating
the variance (\ref{var}) is a perfect chain, in which all $\beta_i$ are
the same, $\beta_i = \bbeta$ and all $\alpha_i$ are zero. In this case, $\tH$ can 
be diagonalized exactly:
\ba
r_{\ell i} & = & C \sin k_\ell (i + 1) \, , \label{re} \\
\tE_\ell & = & 2 \bbeta \cos k_\ell \label{Ee} \, ,
\ea
where $k_\ell$ runs over $n$ integer multiples of $\pi/ (n+1)$, and 
$C = [2 / (n + 1)]^{1/2}$ is a normalization constant.
Then,
\[
r_{\ell, n-1} = C \sin k_\ell n = \pm C \sin k_\ell \, ,
\]
and the variance (\ref{var}) is 
\be
(\Delta E_\ell)^2 =  C^2 ( \bbeta^2 - \tE_\ell^2 / 4 ) \, .
\label{circle}
\ee
We see that $\Delta E$ as a function of $\frac{C\tE}{2}$ is a semicircle of 
radius $\bbeta C \sim  W /\sqrt{n}$.

\begin{figure}
\centering
\includegraphics[width=10cm]{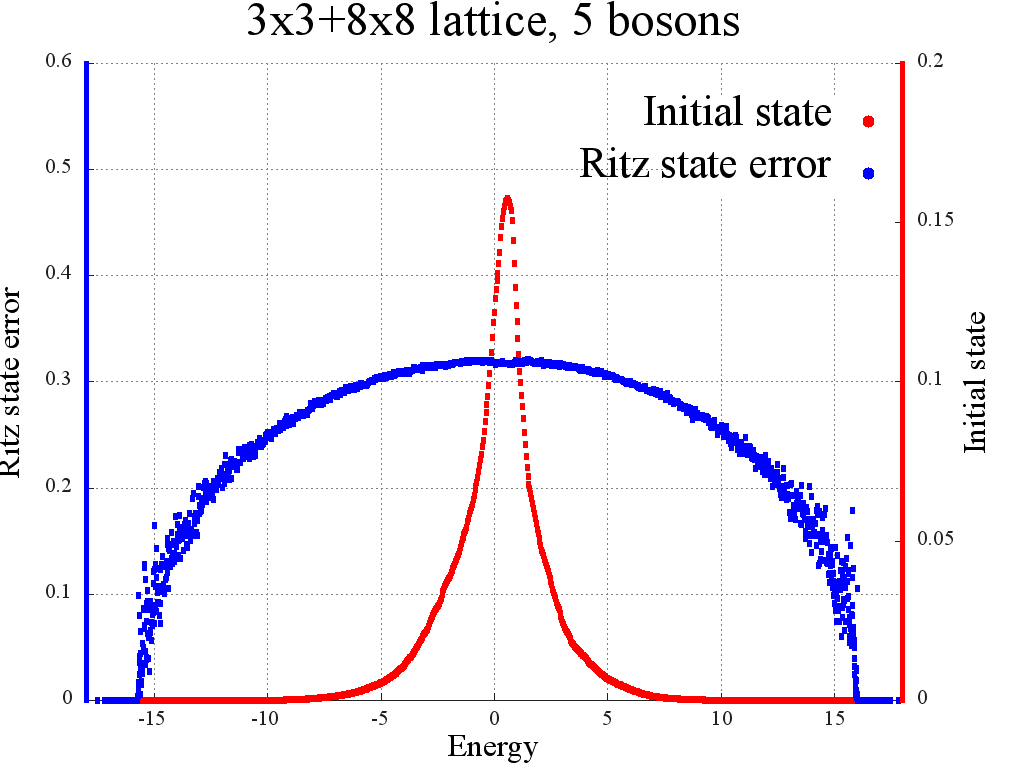}
\caption{After diagonalizing the matrix of the Hamiltonian in the Krylov subspace ($n=1240$), 
we plot, as a function of energy, the error (the standard deviation $\Delta E_l$) in energy 
of each Ritz vector (blue). Some states
at the ends of the spectrum have converged whereas in the middle the error is maximal. 
In red, we plot the square of the energy wave function of the initial 
state $|\bra{\ell}\psi(t=0)\rangle|^2$ as a function of the Ritz eigenvalue $\tE_\ell$. 
The distribution is relatively narrow because the initial state is an eigenstate of 
the Hamiltonian $H_A+H_B$ describing two decoupled boxes ($A$ and $B$),
and the interaction  $H_{AB}$ is small since $A$ and $B$ are connected only 
by two links.}
\label{ErrorInState}
\end{figure}

How much do the small random fluctuations of $\alpha$ and $\beta$ modify this picture?
In one dimension, Anderson localization is very powerful: if the chain were infinite,
arbitrarily small disorder would localize {\em all} the states. In our case, however,
the chain is finite, of length $n$, and disorder is weak, of order $1/\sqrt{N}$. 
For weak disorder, the localization length (in one dimension)
scales as inverse of the disorder potential 
squared \cite{Thouless}. We conclude that at
$n \ll N$ only relatively few states will be localized. Localization first begins in the
part of the spectrum where the density of states in the ideal system 
is the largest: in our case,
at the edges of the spectrum, near $\tE = \pm 2 |\bbeta|$. Indeed, this is precisely 
where the Lanczos recursion first converges. Thus, at $ n \ll N$, we expect that
the semicircle
represented by (\ref{circle}) will remain mostly intact in the presence of disorder, 
except for the largest and smallest eigenvalues, where $\Delta E$ will be close to zero.
This agrees very well with the numerical results, see Fig.~\ref{ErrorInState}.

The decrease of the radius of the circle, as $1/ \sqrt{n}$, with the size of the
Krylov subspace can be taken as an indication that each Ritz vector contains mostly
(i.e., with substantial amplitudes) only those 
eigenstates of the full $H$ that lie in the narrow, of a width of order $W/\sqrt{n}$,
band of energies near $\tE$. We have used that for estimating both the diagonal and
off-diagonal matrix elements of an operator between the Ritz states, eqs.~(\ref{estRitz})
and (\ref{est-off-diag}), respectively.

\section{Entanglement entropy generation}

\begin{figure}
\centering
\subfloat[]{\includegraphics[width=6.5cm]{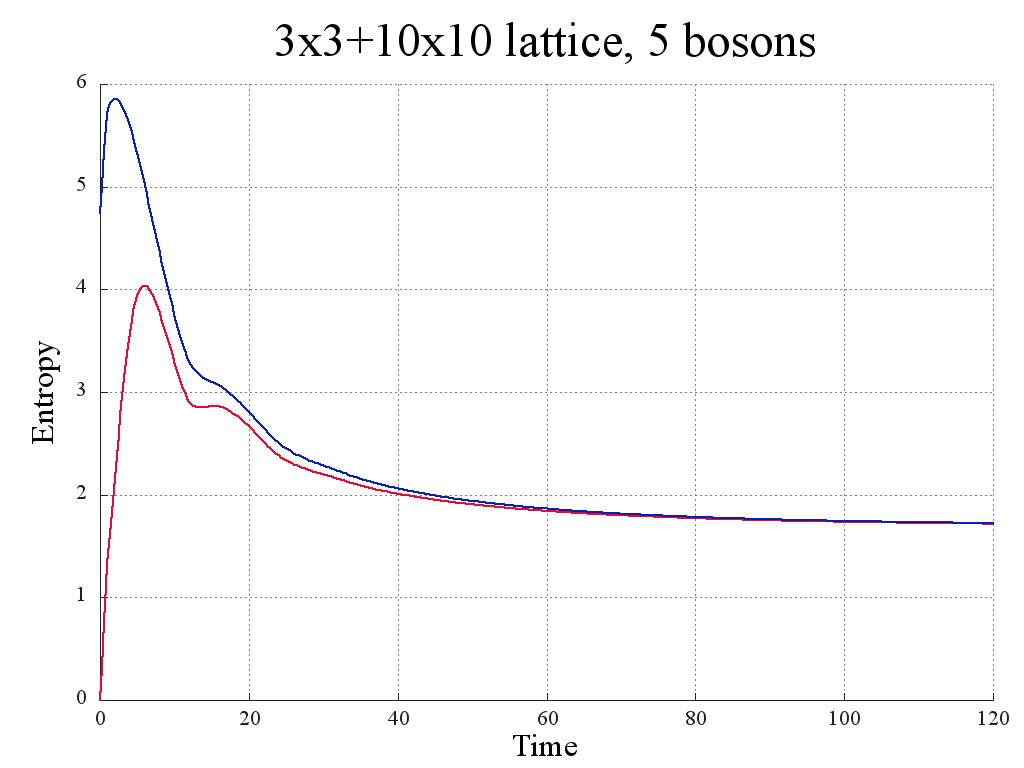}}
\subfloat[]{\includegraphics[width=6.5cm]{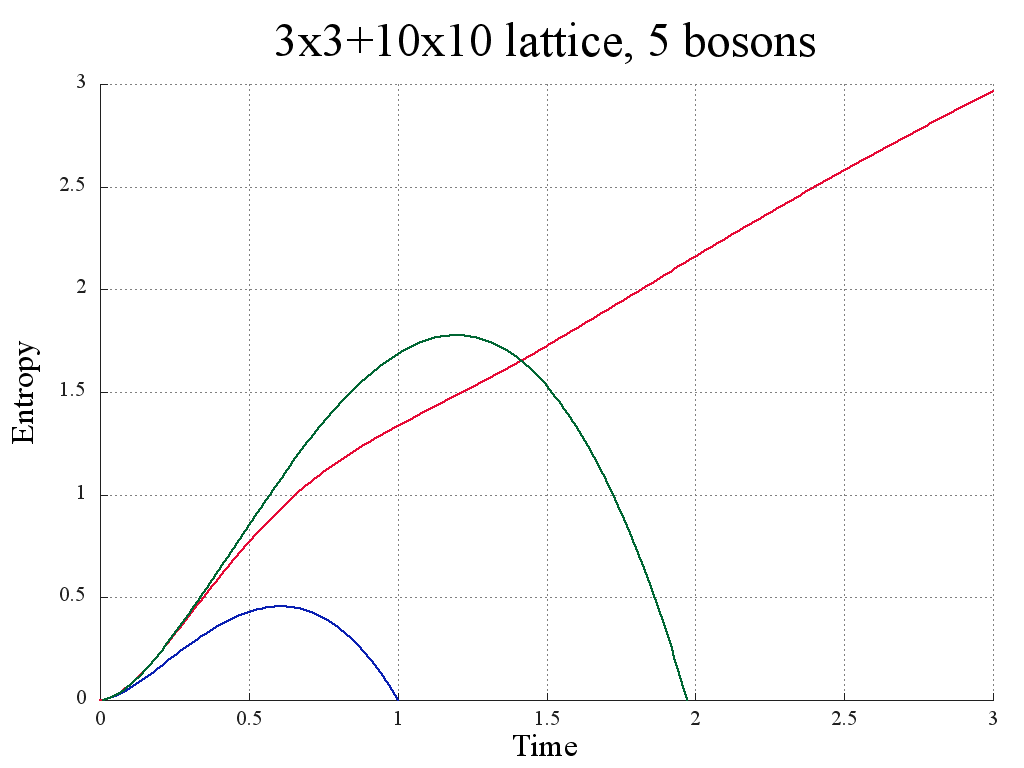}} 
\caption{Entanglement entropy between subsystems $A$ and $B$ as a function of time (red curves). In (a) it is seen how the entropy grows to a maximum and then, being bounded 
by the thermodynamical entropy (blue curve), decreases as the number of particles in subsystem $A$ decreases, reducing the available number of states.  
 The initial growth is depicted in (b). The leading term (blue curve) is $S\simeq - \bra{\psi_0} n_7+n_8 \ket{\psi_0}\, t^2 \ln t^2 =-1.24\, t^2 \ln t^2 $ where
$n_7$, $n_8$ are the initial occupation numbers of the sites of $A$ that are
in contact with $B$. A better approximation is $S\simeq - \bra{\psi_0} n_7+n_8 \ket{\psi_0}\, t^2\ln t^2 + 1.7 t^2$ (green curve), where the 
constant $1.7$ in the subleading behavior is obtained by fitting the numerical data.}
\label{Svst}
\end{figure}

Given the spatial structure of the lattice we have studied, 
it is natural to divide the system into subsystem $A$, the $3\times 3$ block, 
and subsystem $B$, the rest. 
Accordingly, we introduce  
Hamiltonians $H_{AA}$, $H_{BB}$ associated with each region and $H_{AB}$, their interaction. 
They have the same form as $H$ in eq.(\ref{b2}) but with the sum over
indices restricted to the corresponding subregions. Thus
\beq
 H = H_{AA}+H_{BB}+H_{AB} \, . \label{d0}
\eeq
In this section we consider the same initial state as in the 
previous sections, namely, the product state
\beq
 \ket{\psi(t=0)} = \ket{\psi_0} = \ket{\psi_{A0}} \otimes \ket{0_B}  \, , \label{d1}
\eeq
where $\ket{\psi_{A0}}$ is an eigenstate of $H_{AA}$ and $\ket{0_B}$ is the empty state for region $B$. This state evolves in time defining a density matrix for 
subsytem $A$:
\beq
 \rho_A(t) = \tr_B \rho(t) = \tr_B \ket{\psi(t)}\bra{\psi(t)} \, ,   \label{d2}
\eeq
and the entanglement entropy
\beq
S_{AB}(t) = - \tr \rho_A(t) \ln \rho_A(t)     \, .     \label{d3}
\eeq
 Numerically the initial state was taken as the product of an eigenstate of $H_{AA}$ and the empty state of region $B$. The resulting $S_{AB}(t)$ in plotted in fig.\ref{Svst}.

The overall shape of the curve with the entropy rising and then decreasing is similar to that
discussed by Page \cite{Page}, who computed the average
entanglement entropy as function of 
the dimension of a subsystem, under the assumption that the entire isolated system is in 
a random pure state.

Combining the numerical data with analytical estimates, we can understand the 
$S_{AB}(t) \equiv S(t)$ curve in quite a bit of detail. 
The initial growth of the entropy follows the law
\beq
 S = - (\Delta E)^2\, t^2 \ln \left(\frac{t}{t_0}\right)^2   \, ,   \label{d4}
\eeq
where $(\Delta E)^2 = \bra{\psi_0} H^2 \ket{\psi_0}-\bra{\psi_0}H\ket{\psi_0}^2$ is the energy spread of the initial state\footnote{This result depends on the properties of the initial state. Generically the behavior is $S\sim -t^p \ln t^p$ for some integer $p\le 1$.}. After fitting the remaining constant $t_0$ from the data, the curve is plotted in fig.\ref{Svst} showing that, for short times, it is a good fit to the numerical result. 

To derive (\ref{d4}), one can start from the equation for $\rho$:
\beq
 \partial_t \rho = -i [H,\rho]  \, ,    \label{d5}
\eeq
to obtain
\beq
 \rho(t) = \rho(0) -i [H,\rho(0)] t - \half [H [H,\rho(0)]] t^2 + \ldots     \label{d6}
\eeq
 Taking trace over $B$, we observe first that the leading behavior of $\rho_A$ at $t\rightarrow 0$ is
\beq
 \rho_A = \rho_A(0) + t^p \rho_A^{(p)}      \label{d7}
\eeq
 for some integer $p$. Although the density matrix is analytic at $t=0$ the entropy in not necessarily so. The density matrix $\rho_A(0)$ has an eigenvalue $\rho_0=1$ 
corresponding to the initial state; all the other eigenvalues vanish. 
By the usual rules of perturbation theory, the eigenvalues of $\rho_A(t)$ are
\beq
 \rho_0 =1 + t^p \bra{\psi_{A0}} \rho_A^{(p)} \ket{\psi_{A0}} = 1 + t^p \rho_{00} , \ \ \ \  \rho_{a\neq 0} = t^p \rho_{aa}  \, ,   \label{d8}
\eeq
 where $\rho_{00} = \bra{\psi_{A0}} \rho_A^{(p)} \ket{\psi_{A0}}$ and $\rho_{aa}$ are the eigenvalues of $\rho_A^{(p)}$ projected over the subspace orthogonal 
to the initial state. The entropy, to leading order, is given by 
\beq
 S \simeq - \rho_0 \ln \rho_0 - \sum_{a\neq 0} t^p \rho_{aa} \ln (t^p \rho_{aa}) \simeq - t^p \ln(t^p) \, \sum_{a\neq 0} \rho_{aa} + \cO(t^p)   \, .  \label{d9}
\eeq
 Since the correction to the density matrix has zero trace, we conclude that the leading order behavior of the entropy at short times is
\beq
 S \simeq t^p \ln(t^p) \bra{\psi_{A0}} \rho_A^{(p)} \ket{\psi_{A0}}   \, ,    \label{d10}
\eeq 
 where $p$ is the order of the first non-vanishing term in the Taylor expansion of $\rho_A$. If $\bra{\psi_{A0}} \rho_A^{(p)} \ket{\psi_{A0}} =0$  the entropy
 behaves initially as $t^p$, namely, without the logarithmic factor. This is a quite generic result. Let us check that for our system $p=2$ as claimed before. 

 Consider first the linear term
\beq
 \rho_A^{(1)} = -i \tr_B [H,\rho(0)]   \, .    \label{d11}
\eeq 
 The initial state considered is an eigenstate of $H_{AA}+H_{BB}$, therefore
\beqa
 \rho_A^{(1)} &=& -i \tr_B [H_{AB},\rho(0)] = -i \sum_{E_{Bn}} \bra{E_{Bn}}  H_{AB} \ket{0_B}\otimes\ket{\psi_{A0}}\bra{\psi_{A0}}\otimes\bra{0_B}{E_{Bn}}\rangle \non\\
  && + i \sum_{E_{Bn}} \langle{E_{Bn}}  \ket{0_B}\otimes\ket{\psi_{A0}}\bra{\psi_A(0)}\otimes\bra{0_B} H_{AB}\ket{E_{Bn}}   \label{d12}\\
  &=& -i \bra{0_B}  H_{AB} \ket{0_B} \ket{\psi_{A0}}\bra{\psi_{A0}}  + i \ket{\psi_{A0}}\bra{\psi_{A0}} \bra{0_B} H_{AB}\ket{0_B} \, . \nonumber
\eeqa
This vanishes since, for the particular $H_{AB}$ we are considering $\bra{0_B}H_{AB}\ket{0_B}=0$. Thus, we are left to consider the second order term. 
If it does not vanish, then, as follows from eq.(\ref{d10}), we only need its mean value in the initial $A$ state:
\beqa
\bra{\psi_{A0}} \rho_A^{(2)} \ket{\psi_{A0}} &=& - \half \bra{\psi_{A0}}\ \tr_B [H [H,\rho(0)]]\ \ket{\psi_{A0}}  \label{d13}\\
     &=& - \half  \tr \rho_A(0) [H [H,\rho(0)]] \, ,\label{d14}
\eeqa
where, by a slight abuse of notation, $\rho_A(0)$ is taken as the operator that projects the state of subsystem $A$ onto $\ket{\psi_{A0}}$ and acts as the identity on $B$. 
 Following that notation, we find that
\beq
 \rho(0) \rho_A(0) = \left(\ket{\psi_{A0}}\otimes\ket{0_B}\bra{0_B}\bra{\psi_{A0}}\right) \left(\ket{\psi_{A0}}\bra{\psi_{A0}}\right) = \rho(0) \, ,
\eeq
and  also that 
\beq
\bra{0_B}\bra{\psi_{A0}} H_{AB} \ket{\psi_{A0}} \ket{E_{Bn}} =0 \, ,
\eeq 
 where the last result depends on $\ket{\psi_A(0)}$ being an eigenstate of the total occupation number. It follows that
\beq
 \bra{\psi_{A0}} \rho_A^{(2)} \ket{\psi_{A0}} = - \bra{\psi_0} H^2 \ket{\psi_0} + \bra{\psi_0} H\ket{\psi_0}^2 = - (\Delta E)^2  \, , \label{d15}
\eeq
where $\Delta E$ is the dispersion in energy of the initial state.  Thus
\beq
 S_{AB}(t) \simeq - t^2 (\Delta E)^2 \ln \frac{t^2}{t_0^2}  \, ,  \label{d16}
\eeq
 where the time $t_0$ determines the subleading $t^2$ term. An interesting consequence of
(\ref{d16}) is that the initial growth of entropy is directly related to the spread
in energy of the initial state. In our case, after some algebra we find
\beq
 \Delta E_0^2 = \bra{\psi_0} H_{AB} H_{AB} \ket{\psi_0} = J_1^2  \bra{\psi_0} \sum_i n_i \ket{\psi_0}   \, , \label{d17}
\eeq
 where $J_1$ is the hopping amplitude in (\ref{b2}), and $n_i$ are the occupation numbers of all sites of subsystem $A$ 
that are in contact with subsystem $B$: in the present case $i=7,8$. To derive this result it is necessary
that the sublattice $B$ is initially empty. Finally, for the initial growth we obtain
\beq
  S_{AB}(t) \simeq - t^2 J_1^2 \bra{\psi_0} n_7+n_8 \ket{\psi_0}  \ln \frac{t^2}{t_0^2}  \, ,   \label{d18}
\eeq 
 which was used to fit the curve in fig.\ref{Svst}. 
From the physical point of view, it is interesting to note that 
this initial growth is due to streaming of particles 
from the small box into the vacuum of region $B$.

The growth of entropy cannot continue forever, as there is a maximum entropy for a density matrix with 
given mean values of energy and particle number. It is given by the thermodynamical entropy associated with the thermal density matrix. This is an exact result
valid for any system, small or large: the entropy will always be bound by the 
thermodynamical value. Now, when the bosons start to leave the small box, at some point 
the available
number of states decreases and so does the thermodynamical entropy. 
Therefore, quite generically, the entanglement entropy should also start to decrease. 
This is seen in 
fig.\ref{Svst},  where the thermodynamical entropy is plotted and shown to be an upper bound for the entanglement entropy. A surprising result from the numerics is that 
the entanglement entropy is quite close to the thermodynamical entropy much earlier than the thermalization time. This implies that the small subsystem can thermalize 
by streaming particles into vacuum. Notice that here we mean actual thermalization, 
where the subsystem is in a mixed state close to thermodynamical equilibrium. 
This result, though, 
depends on the initial state having a relatively
large energy. If we start from a low-energy state, 
the entropy rises until the whole system reaches equilibrium. 

\subsection{ETH property of the entropy}

\begin{figure}
\centering
\subfloat[$N=5984, n=1240$    ]{\includegraphics[width=6.5cm]{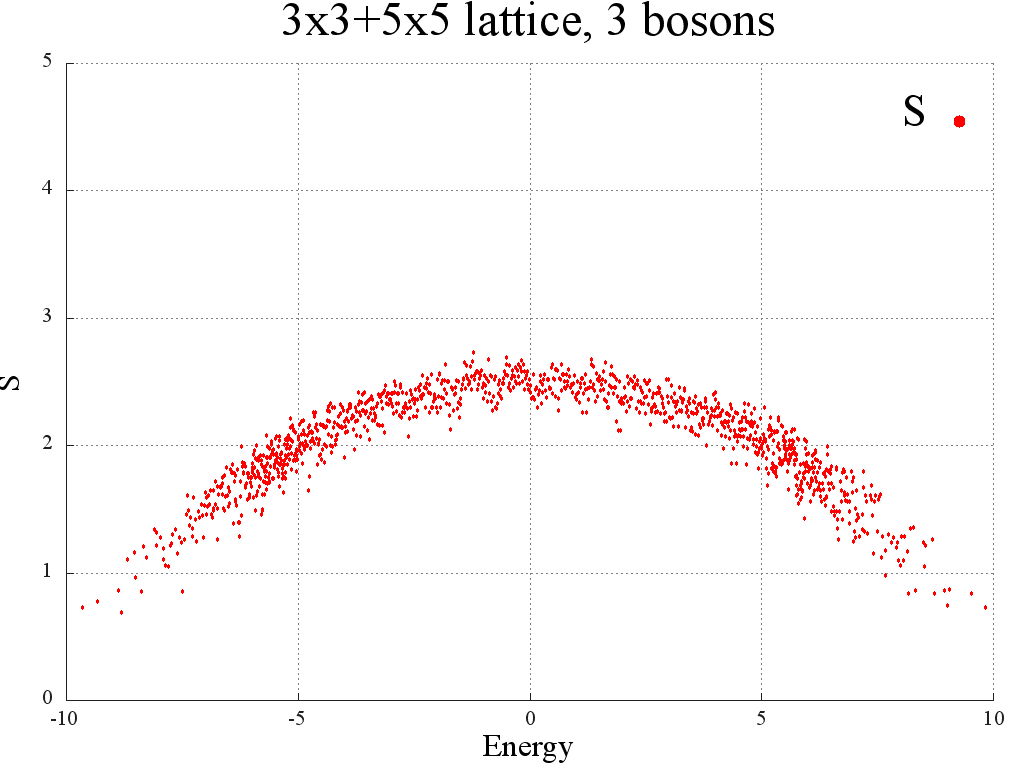}}
\subfloat[$N=53130, n=1240$   ]{\includegraphics[width=6.5cm]{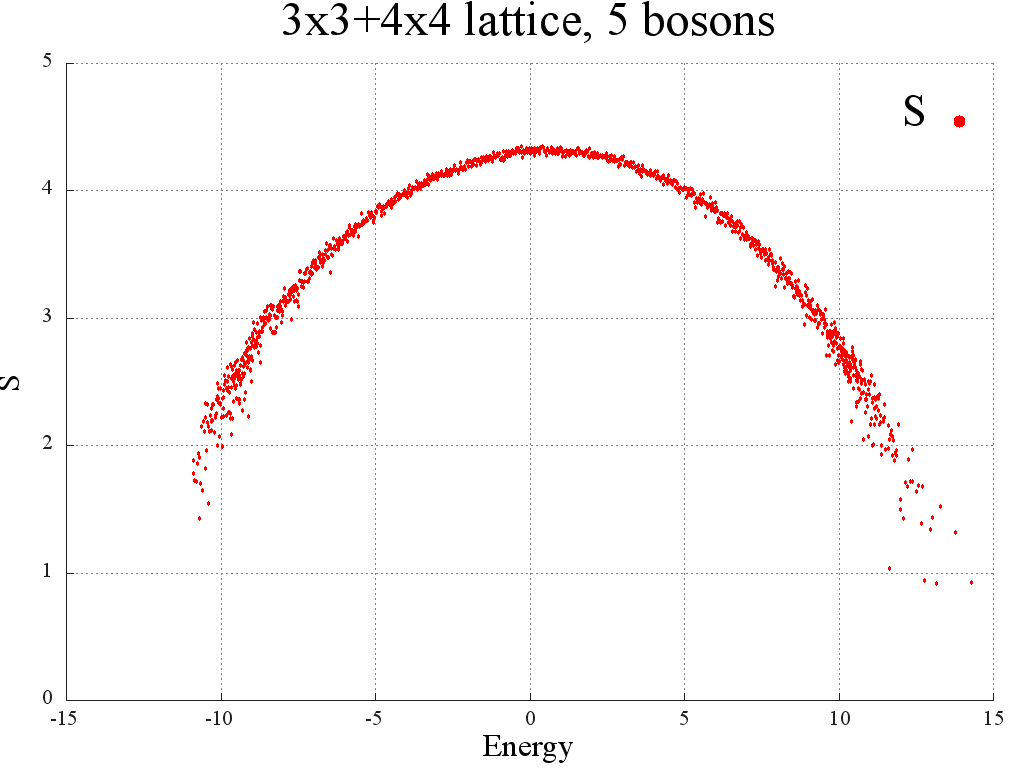}} \\
\subfloat[$N=1221759, n=1240$ ]{\includegraphics[width=6.5cm]{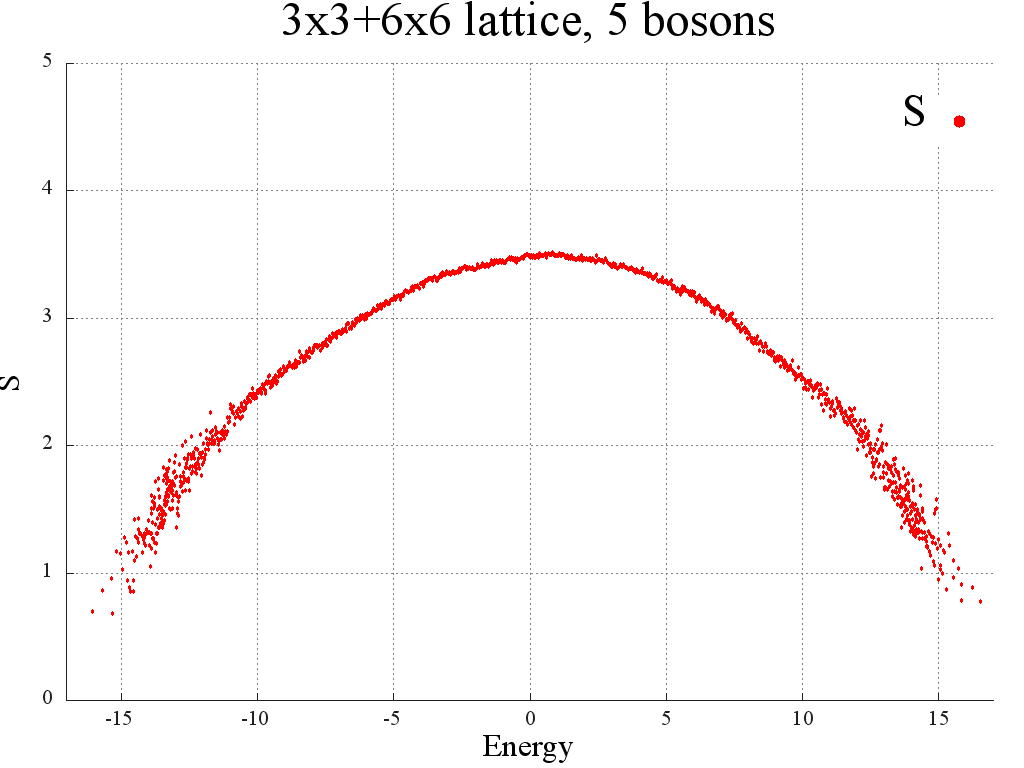}}
\subfloat[$N=15020334, n=1240$]{\includegraphics[width=6.5cm]{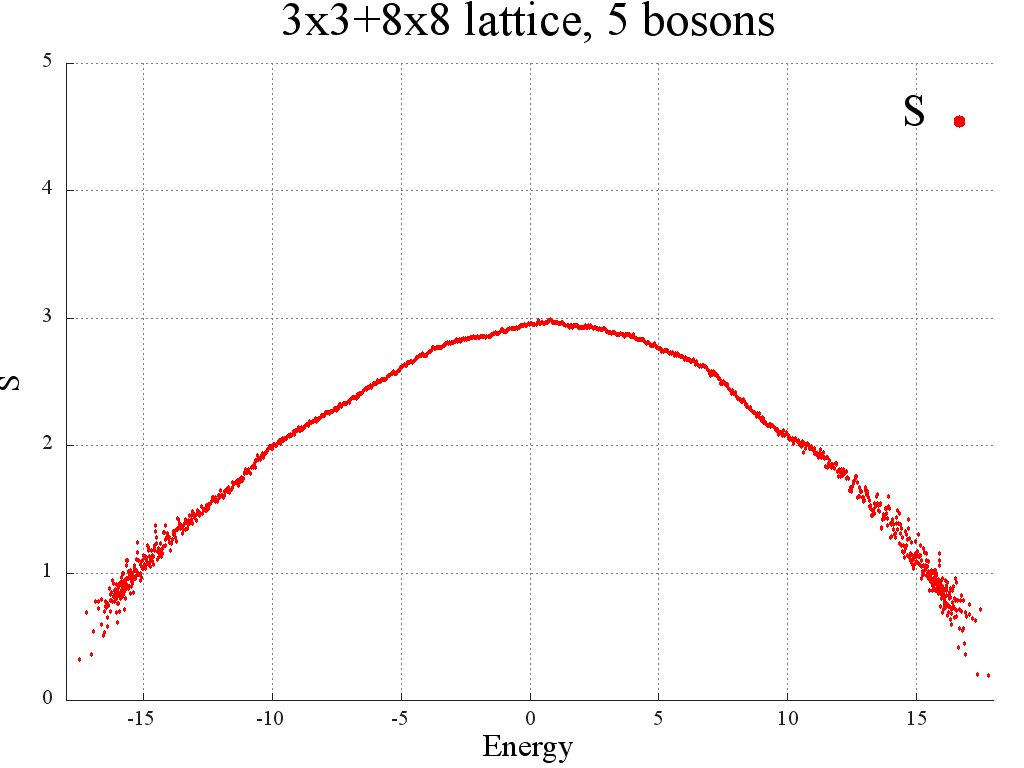}}
\caption{Entanglement entropy $S_{AB}$ of the subsystems $A$ and $B$ computed in the Ritz states and plotted as a function of energy. 
It is clearly seen that the function becomes smooth as $N$ becomes larger.}
\label{SETH}
\end{figure}

 As it was discussed previously in this section, the entanglement entropy thermalizes in the same way as the occupation number. It is therefore interesting to plot the
value of the entanglement entropy for the subsystem in each of the Ritz states and see if it becomes a smooth function of the energy as the system grows larger.  
This is done in fig.\ref{SETH} where it is seen that, as was the case for the occupation 
number, the entropy does indeed become a smooth function. 

\section{Conclusions} 
\label{conclusions}

 In this paper we studied, numerically, a particular system that displays thermalization behavior
while being small enough to make a simulation of its quantum evolution feasible. 
Although this system may be
small by thermodynamical standards,  we still consider cases where the Hilbert space has dimension $\gtrsim 10^8$. The reason why that could be done is that
the time evolution to thermalization time $t_{th}$ occurs in a subspace of dimension $t_{th} \Delta E \sim 10^3$ where $\Delta E$ is the spread in energy of the 
initial state. This is the Krylov subspace associated with the initial state. Because construction of
the Krylov subspace requires applying the full Hamiltonian, we are still subject to the above mentioned 
practical restriction on the dimension $\sim 10^8$. 

For thermalization to occur, it is sufficient that the Eigenstate Thermalization Hypothesis (ETH)
is valid in the Krylov subspace only. Our numerical study shows that it is indeed valid there and,
for a Krylov subspace of a fixed dimension, holds better and better as 
we increase the dimension of the full Hilbert space. This is an important test of ETH.
 
While we can understand thermalization for our system as a result of the ETH in the Krylov
subspace, the question remains if it is valid there because it is valid in the
whole space or because the  eigenstates of the projected Hamiltonian (the Ritz vectors) 
represent averaged properties of 
the underlying exact eigenstates, resulting in the averages of various operators being smooth. 
 Numerically we cannot answer this question. In the present context, answering it would be equivalent 
to testing all possible initial states. It seems plausible that the ETH extends to the
whole Hilbert space but it is also possible that only a subset of states thermalize. In that case the
ETH would be valid only in the Krylov subspaces associated with those
states. 
 
A related aspect of the calculation was a study of how two initially independent regions 
become entangled as a result of the evolution and as measured by 
the entanglement entropy $S_{AB}(t)$ as a function of time.  We found that the growth of $S_{AB}$ is
initially of the form $S_{AB}(t)\simeq -(\Delta E)^2 t^2 \ln \frac{t^2}{t_0^2}$ and then becomes approximately
linear until $S_{AB}$ reaches the maximum allowed, namely the thermodynamic entropy, afterwards it begins to decrease. 
All this happens as a result of streaming of particles into vacuum. When vacuum is not there anymore,
i.e., the larger container fills up, the entanglement entropy decreases slower. This continues 
until the full system 
reaches thermal equilibrium, in the sense that the mean occupation numbers of all lattice sites are 
constant, up to small fluctuations.
After that, the entanglement entropy of the subsystem remains constant and 
equal to the thermodynamical one. There is an analogy between this process and the formation and evaporation of a black hole. In that case the entanglement entropy between the black hole and the Hawking radiation has a similar behavior. 

Regarding quantum black holes, the ETH  implies that, for certain ``thermal'' operators, the expectation values
in energy eigenstates depend only on the total energy and therefore are the same for all the back hole microstates 
with close-by energies. 
The metric appears to be one such operator. Indeed, the no hair theorem of classical gravity
says that the (outside) metric is completely determined by the black hole mass (and other conserved 
quantities such as angular momentum or charge). The ETH, as applied to the metric, would be a quantum
version of this statement. 

If the ETH applies to the metric, the latter should have the same value 
if computed in any arbitrary microstate or in a thermal density matrix. As such, 
it contains no information whatsoever on the nature of the microstate. 
Notice that this point of view is different from the perhaps more conventional one, according to which 
a given microstate has no well 
defined metric. In that view, the metric is ``fuzzy'' in the individual energy eigenstates, 
and one needs
to craft special coherent superpositions of them to obtain a well-defined classical metric.
If all microstates indeed have the same metric, it is meaningless, for example, to ask if the information on 
the microstate is localized near the horizon or at the 
singularity, simply because no microscopic information is contained in the metric. 
The same would be true for the Hawking radiation: insofar as it is computed solely from the black hole metric 
it cannot contain any microscopic information. Its properties should therefore be described by operators that
are ``thermal''in the ETH sense.
  
 Since our very notion of locality is based on the metric, it is possible that locality is an emergent, 
as opposed to fundamental, property of quantum gravity, similar to the second law of thermodynamics in ordinary statistical mechanics. 
 This would imply that various properties of black holes seen in classical gravity, for example, the
impossibility of leaving the black hole interior and perhaps even the speed of light limit, are statistical 
laws only, which, for large black holes, hold with an overwhelming probability but still not absolutely.
Trying to describe collapse of matter to a black hole by using the metric alone is equivalent to describing,
in the present context, the expansion of a lattice gas by computing the mean
occupation number as a function of time. It is a good description if only thermodynamical or average information 
is desired. The information about the initial state is lost.  

In summary, the ETH appears to provide a good starting point for addressing standard but still unanswered questions 
concerning the properties of black holes in quantum gravity.  

%In application to quantum black holes, the ETH would suggest that the
%properties seen in classical gravity, such as 
%the classical metric itself
%or the Hawking radiation, are ``local'' or ``thermal'' and therefore can be computed 
%with equal results in an eigenstate of energy
%or in a thermal density matrix. As a consequence, they do not contain any information about 
%the particular microstate the black hole is in. Since our notion of locality is based on
%the space-time metric, it may also be that 
%the question as to where in space-time the information about the microstate is localized 
%is meaningless. In other words, locality could be an emergent,
%as opposed to fundamental, property of quantum gravity, similar to the
%second law of thermodynamics in ordinary statistical systems. 

\section{Acknowledgments}

 We are very grateful to Peter Ouyang for discussions and collaboration in the initial stages of this work. 
We are also grateful to M. Srednicki for various comments and suggestions along the way and to M. Rigol for
suggestions on how to strengthen the numerical results. 

 This work was supported in part by the DOE through grant \protect{DE-SC0007884}. 
In addition, the work of M.K. was partially supported by the NSF through a CAREER Award PHY-0952630. 
M.K. also wants to thank the hospitality of the Perimeter Institute and the KITP (Santa Barbara) while 
part of this work was being done.

%\section{Appendix}

\end{document}